\documentclass[11pt,reqno]{article}
\usepackage[lmargin=2cm, rmargin=2cm]{geometry}
\usepackage[dvips]{graphicx}
\usepackage{amssymb}
\usepackage{appendix}
\usepackage[utf8]{inputenc}
\usepackage[english]{babel}
\usepackage{amsmath, amssymb, graphics}
\usepackage{amsfonts, color}
\usepackage{mathrsfs}

\allowdisplaybreaks \numberwithin{equation}{section}

\newcommand{\mathsym}[1]{{}}

\newcommand{\g}{\gamma}

\newcommand{\mc}{\mathcal}

\newcommand{\p}{\partial}

\newcommand{\rar}{\rightarrow}

\newcommand{\nn}{\nonumber}
\newcommand{\im}{\mathrm{Im}}
\newcommand{\re}{\mathrm{Re}}

\def\bfone{\relax{\rm 1\kern-.35em 1}}

\makeatletter \@addtoreset{equation}{section} \makeatother

\usepackage{hyperref}

 \allowdisplaybreaks[3]

\title{\vspace{-0.5cm}\textbf{ Electric non-extremal solution in AdS }}
\date{\today}

\begin{document}

\begin{titlepage}
 \thispagestyle{empty}
 \begin{flushright}
     \hfill{ITP-UU-14/16 }\\
     \hfill{SPIN-14/14 }\\
 \end{flushright}

 \vspace{80pt}

 \begin{center}
     { \LARGE{\bf      {First order flow for non-extremal AdS black holes\\ [5mm] and mass from holographic renormalization }}}

     \vspace{60pt}

Alessandra Gnecchi and Chiara Toldo \\[8mm]
{\small\slshape
Institute for Theoretical Physics \emph{and} Spinoza Institute, \\
Utrecht University, 3508 TD Utrecht, The Netherlands \\

\vspace{5mm}

{\upshape\ttfamily A.Gnecchi@uu.nl,  C.Toldo@uu.nl}\\[3mm]}

\vspace{8mm}

     \vspace{20pt}

    \vspace{20pt}

     {\bf Abstract}
     \end{center}
In this paper we present a first order formulation for non-extremal Anti-de Sitter black
hole solutions in four dimensional $\mathcal{N}=2$ U(1)-gauged Supergravity. The dynamics
is determined in terms of a quantity $\mathcal{W}$ which plays the role of a
superpotential for the gauging potential in the action. We show how
the first order flow arises from writing the action as a sum of squares and we identify the superpotential driving the first order flow for two classes of solutions  (electric and magnetic) of the $t^3$ model.
After identifying $\mathcal{W}$, we study the Hamilton-Jacobi holographic
renormalization procedure in presence of mixed boundary conditions for the scalar fields. We compute the renormalized on-shell action and the mass of the black hole configurations. The expression obtained for the mass satisfies the first law of thermodynamics.

 \vspace{10pt}
\noindent 

\end{titlepage}


\thispagestyle{plain}

\tableofcontents

\baselineskip 6 mm

\section{Introduction}

Supersymmetry plays a fundamental role in string theory and supergravity. It also provides us with a very powerful tool to find new solutions in such theories. BPS configurations can be found by solving first order flow equations that arise from the preservation of some amount of supersymmetry. 

For asymptotically flat black holes in four-dimensional $\mathcal{N}=2$ supergravity, the dynamics is determined  by an effective black hole potential $V_{BH}$, function of the scalar fields and the electromagnetic charges (for a review see \cite{Andrianopoli:2006ub}) 
\begin{equation}
V_{BH} (z, \bar{z}, p^{\Lambda},q_{\Lambda})= g^{i \bar{\jmath}} D_i Z \bar{D}_{\bar{\jmath}} \bar{Z} + |Z|^2 \,,
\end{equation}
where $Z$ is the central charge of the theory \cite{Gibbons:1996af,Ferrara:1995ih,Ferrara:1997tw}, which plays the role of a  superpotential driving the BPS flow. In the case of non-BPS configurations one can still find a real function  $W$  satisfying 
\begin{equation}\label{vblack}
V_{BH} = 4 g^{i \bar{\jmath}}  \partial_i W \partial_{\bar{\jmath}} W + W^2 \,,
\end{equation}
playing the role of a "fake superpotential" \cite{Ceresole:2007wx,Andrianopoli:2007gt,Galli:2010mg}.

In analogy with the flat case, also for Anti-de Sitter (AdS) configurations a first order flow has been found for BPS \cite{Cacciatori:2009iz,Dall'Agata:2010gj,Hristov:2010ri,Katmadas:2014faa} and extremal non-BPS \cite{Klemm:2012vm,Gnecchi:2012kb} black holes in $U(1)$-gauged supergravity. A superpotential has been found also in this setup, however it does not satisfy any relation of the form \eqref{vblack}.

First order flow equations for non-extremal configurations are harder to find. In general thermal configurations requires one to solve the full system of Einstein-Maxwell-scalar equations of motion. For asymptotically flat black holes a first order formulation for non-extremal solutions has been related to the Hamilton-Jacobi formalism \cite{Perz:2008kh,Trigiante:2012eb}.

Motivated by interest in holographic applications, we investigate such formalism for non-extremal AdS black holes. For instance, bound states of charged AdS black holes in $\mathcal{N}=2$ supergravity have recently been used in the holographic study of glassy systems \cite{Anninos:2013mfa}. Having a first-order formulation would facilitate the task of finding new solutions and might shed light on some open problems concerning the relation between the moduli and the entropy of non-extremal black holes, or the existence of multicenter AdS solutions \cite{Anninos:2013mfa,Chimento:2013pka}.

In this paper we find a first order formalism for non-extremal four-dimensional Anti-de Sitter black holes and we present the corresponding equations for the warp factors and the scalar fields. Previous studies on this topic, in addition to those already mentioned, can be found in \cite{Trigiante:2012eb}\cite{Miller:2006ay}-\nocite{Lu:2003iv,Elvang:2007ba}\cite{Cardoso:2008gm}, and for black branes in \cite{Barisch:2011ui,Barisch-Dick:2013xga}.

We work in the framework of $\mathcal{N} =2$ $U(1)$-gauged supergravity in four dimensions. Inspired by the work of \cite{Cardoso:2008gm} in five dimensions, we derive a first order flow by rewriting the one-dimensional effective Lagrangian as a sum of squares plus a term whose variation vanishes when the fields satisfy the first order equations. Because of this non-squared term one cannot achieve a fully-BPS rewriting, indicating that the first order flow will correspond in general to non-extremal (thus non-BPS) configurations.
The first order equations we obtain, along with a Hamiltonian constraint on the charges, are sufficient to solve the full system of equations of motion of the original action. 

The squaring procedure is valid for Very Special geometries in absence of axions. We derive it in two different cases, namely when the black hole charges are electric and magnetic.

It is important to notice that the flow is driven by a quantity called "superpotential", that is related to the gauging potential of our supergravity theory by the following relation:
\begin{equation}\label{eqpot0}
V_g (\phi) = g^{ij} \frac{\partial W}{ \partial \phi^i} \frac{\partial W}{\partial \phi^j} -3 W^2\,.
\end{equation}
Notice that the superpotential $W$ is only related to the scalar potential of the gauging.  The black hole electromagnetic charges are only required to satisfy a Hamiltonian constraint involving $V_{BH}$. Remarkably, the first order equations we found are analogous of those obeyed by (uncharged) domain wall solutions in AdS \cite{Cardoso:2008gm}.

As an explicit example, we re-derive the solutions of \cite{Duff:1999gh} and  \cite{Klemm:2012yg,Toldo:2012ec} by means of the first order equations and we identify their corresponding superpotential $W$. We finally comment on the BPS limit of this flow.

The second part of this paper is devoted to the computation of the black hole mass for Anti-de Sitter configurations.  Defining the mass in AdS  is usually  nontrivial task, due to the fact that the Komar integral is divergent. Holographic renormalization techniques (see for example \cite{deBoer:1999xf,Bianchi:2001kw,deHaro:2000xn,Papadimitriou:2005ii}) remove the divergencies of the boundary stress-energy tensor by adding additional surface terms to the bulk theory action. These counterterms are built out of curvature invariants of a regularized boundary (which is sent to infinity after the integration) and thus they do not alter the bulk equations of motion.
This yields a well-defined boundary stress tensor and a finite action and mass of the system. 

We compute the mass for black hole configurations known in the literature (see \cite{Duff:1999gh,Klemm:2012yg,Toldo:2012ec}) by means of the Hamilton-Jacobi (HJ) holographic renormalization formalism \cite{deBoer:1999xf}. An analysis in this direction has been carried out in \cite{Liu:2004it,Batrachenko:2004fd} for electric AdS black holes.

The black hole solutions under investigation satisfy mixed boundary conditions for the scalar fields, hence they correspond to a multi-trace deformation of the dual field theory. The HJ renormalization  procedure requires the identification of the superpotential $W$ mentioned before and further care is required due to the presence of mixed boundary conditions for the scalar fields. Indeed in this specific case one needs to take into account further finite boundary terms \cite{Papadimitriou:2007sj}.

The formula obtained for the mass satisfies the first law of thermodynamics and coincide with the value obtained with the Astekar-Magnon-Das (AMD) formalism \cite{Ashtekar:1984zz,Ashtekar:1999jx}. Furthermore, we compute the renormalized on shell action and we find that it coincides with the free energy found by integrating the first law.

\section{\label{sec-squaring}Squaring the action of 4d $U(1)$-gauged Supergravity}

Supergravity black holes in asymptotically Anti de Sitter can be studied in a simple setup.

In this paper we consider the case of $\mc N=2$, $U(1)$-gauged  (Fayet-Iliopoulos) Supergravity  coupled to $n_V$ vector multiplets, along the lines of \cite{Cacciatori:2009iz}. The gauged isometry of this theory is an abelian R-symmetry and precisely a $U(1)_g\subset SU(2)_{R}$ group. The only effect of the gauging on the bosonic sector is to introduce in the Lagrangian a potential $V_g$  as \cite{Dall'Agata:2010gj}
\begin{eqnarray}\label{Vgdef}
V_g=-3|\mc L|^2+g^{i\bar\jmath}\p_i\mc L\p_{\bar\jmath}\bar{\mc L}\ ,\qquad \mc L=\langle \mc G, \mc V \rangle\ ,
\end{eqnarray}
where the symplectic vector $\mc G=(g^{\Lambda},g_{\Lambda})^T$ specifies the gauging and $\mc V=(L^{\Lambda}(z,\bar z),M_{\Lambda}(z, \bar z))^T$ are the symplectic sections of $\mc N=2$ special geometry normalized as $M_{\Lambda}\bar L^{\Lambda}-L^{\Lambda}\bar M_{\Lambda}=-i$. The indices are i,$\bar\jmath=1,...,n_V$, and $\Lambda=0,1,..,n_V$.

The gauging also affects the fermions which acquire a charge under $U(1)_g$. This is crucial in the study of supersymmetric solutions, since the BPS equations are modified with respect to the ungauged theory. However, the non extremal Einstein, Maxwell and scalar equations of motions decouple from the fermionic sector and one can neglect the effect of the gauging on the fermions in the construction of black hole solutions.

We will only consider very special geometries, i.e. theories of $\mc N=2$ Supergravity coupled to vector multiplets whose scalars non-linear sigma model is specified by a symmetric rank-3 tensor $d_{ijk}$  \cite{deWit:1992wf}. In particular, one can choose a symplectic frame such that the prepotential of the scalar manifold is 
\begin{eqnarray}\label{sympl-frame}
F(X^{\Lambda})&=&-\frac{i}{4}\sqrt{X^0\hat d^{ijk}X^iX^jX^k}\ ,
\end{eqnarray}
where the hatted tensor is a constant tensor satisfying 
$
\hat d^{ijk}d_{j(lm}d_{np)k}=\frac{64}{27}\delta^i_{(l}d_{mnp)}
$ \cite{Gnecchi:2013mta}.

All very special geometries descend from a 5-dimensional $\mc N=2$ theory coupled to $n_V-1$ vector multiplets. We henceforth only consider the case in which the axions are set to zero. This is consistent with the requirement that the four dimensional scalars are real, and, in the symplectic frame \eqref{sympl-frame}, this implies that the symplectic sections $L^{\Lambda}$ are real, and thus the $M_{\Lambda}$ are purely imaginary (we work out the real-special geometry relations pertaining to this truncation in Appendix \ref{AppA}).

In this setup we consider black holes with  purely electric $\mc Q=(0,q_{\Lambda})$ or purely magnetic $\mc Q=(p^{\Lambda},0)$ charges, in asymptotic $AdS_4$ spacetime supported by purely electric gauging: $\mc G=(0,g_{\Lambda})$.

We derive a first order flow for non-extremal solutions in the general case of symmetric $d_{ijk}$ tensor. As a concrete example, we solve for electric and magnetic black holes in the $t^3$ model with prepotential $F=-2i\sqrt{X^0(X^1)^3}$, that can be embedded in $\mc N=8$ $SO(8)$ gauged supergravity \cite{Cvetic:1999xp}.

\subsection{Setup and conventions}

We consider a generic bosonic action for gravity coupled to a set of $n_s$ scalar and $n_f$ vector
fields given in the form
\begin{eqnarray}\label{action-CK}
S_{4d}=  \int\,d^4x\sqrt{-g} \left( \frac R2 + g_{ij}(z) \p_{\mu} z^i\p^{\mu} z^j +  \mathcal{I}_{\Lambda\Sigma}(z) F^{\Lambda}_{\mu \nu}F^{\Sigma\,\mu \nu}-V_g  \right)\,,
\end{eqnarray}
where $z^i$, $i=0,1,...n_s-1$ are real scalars, $F_{\mu \nu}^{\Lambda} = \partial_{[\mu}A_{\nu]}$ are the field strengths for the vector fields ($\Lambda,\Sigma =0,1,..n_f-1$), and $V_g$ is the scalar potential. 
We assume that the potential can  be written as 
\begin{equation}\label{eqpot}
V_{g}=g^2 \left(-3\mathcal{W}^2+g^{ij}\p_i \mathcal{W}\p_j\mathcal{W} \right)
\end{equation}
and we find this to be true for the examples we treat here.  For instance, all solutions of the model $F = \sqrt{X^0 (X^1)^3}$ which have vanishing axions, the bosonic action therefore can be cast in this form \eqref{action-CK}, with just one real scalar field $z$ and $\Lambda=0,1$:
 \begin{equation}\label{SKpotential}
V_g = - g^2 \left( \frac{\xi_0 \xi_1}{ \sqrt{z}} +\frac{\xi_1^2}{3 }\sqrt{z} \right), \qquad g_{ij}= g_{z z}=\frac{3}{16z^2}\,, \qquad  \mathcal{I}_{\Lambda\Sigma}=\left(\begin{array}{cc}
-z^{3/2}&0\\0&-\frac{3}{\sqrt{z}}
\end{array}\right)\ .
\end{equation}
We do not specify a superpotential $\mathcal{W}$ yet.

\vspace{2mm}

Our procedure of the squaring of the action is however more general, namely we do not need to assume the form of the prepotential. In addition to the usual assumption of staticity and spherical symmetry, we furthermore assume that the sections $L^{\Lambda}$, and therefore the scalars, are real (no axions) and that $Re(\mathcal{N}) =0$, necessary if we want the supersymmetric Lagrangian to fit in \eqref{action-CK}.

Static and spherically symmetric black hole configurations can be cast in this form:
\begin{equation}\label{formaa_metrica}
ds^2 = U^2(r) dt^2 - \frac{dr^2}{U^2(r)} - h^2(r) (d\theta^2 + \sin^2 \theta d \phi^2)\,,
\end{equation}
with
\begin{equation}\label{formuf}
U^2 = e^{K} f(r)\,, \qquad h^2 = e^{-K}r^2\,,
\end{equation}
where for the moment we leave the functions $K(r)$ and $f(r)$ unspecified. Furthermore, the real scalar fields $z^i$ depend just on the radial coordinate $z^i= z^i(r)$, and the Maxwell's and Bianchi equations are solved by
\begin{equation}\label{ans_vec}
F_{tr}^{\Lambda}=\frac{1}{2h^2(r)} \mc I^{\Lambda \Sigma} q_{\Sigma} \,, \qquad
F_{\theta \varphi}^{\Lambda}= \frac12 p^{\Lambda} \sin \theta \,.
\end{equation}

\subsection{Electric configuration}

At this point we consider electrically charged solutions $p^{\Lambda} =0$ with line element \eqref{formaa_metrica}-\eqref{formuf}, where $\mathbf{\widetilde{g}}$ is defined as 
\begin{equation}
\mathbf{\widetilde{g}} = g \tilde \xi
\end{equation}
where $\tilde \xi$ is for the moment an unspecified real constant. The function $f(r)$ appearing in \eqref{formuf} is of this form:
\begin{eqnarray}\label{sol-param1}
f(r)&=&\kappa+\frac{c_1}{r}+\frac{c_2}{r^2}+\mathbf{\widetilde{g}}^2  r^2e^{-2K(r)}\,.
\end{eqnarray}
and the field strengths are purely electric:
\begin{equation}
F_{tr}^{\Lambda}=\frac{1}{2h^2(r)}\mc I^{\Lambda \Sigma}  q_{\Sigma} \,, \qquad
F_{\theta \varphi}^{\Lambda}= 0\,.
\end{equation}

It turns out that we are able to identify first order equations for the warp factor $K(r)$  and the scalar fields $z^i(r)$ in function of the superpotential $\mathcal{W}$ thanks to a suitable squaring of the action. To do this we plug the ansatz \eqref{formaa_metrica}-\eqref{ans_vec} in the action \eqref{action-CK}, and we rewrite the action as a sum of squares, as performed in  \cite{Cardoso:2008gm} for non-extremal five-dimensional gauged supergravity black hole solutions.

We find it useful to divide the action in terms of $S_2$, containing factors of $\mathbf{\widetilde{g}}^2$, and  $S_0$, with zero powers of $\mathbf{\widetilde{g}}$, the gauge coupling constant. Terms in $\mathbf{\widetilde{g}}^1$ are absent.
\begin{equation}
S= S_0 +S_2 
\end{equation}
 In the following, $'$ denotes differentiation with respect to the radial variable $r$ and $\eta_{ab}$ is the 2-dimensional space of constant curvature. We are mostly interested in the black hole examples (spherical horizon topology), namely the case with $\kappa=1$ and $\sqrt{\eta} = \sin\theta d\theta d\phi $. Nevertheless, we keep $\kappa$ unconstrained for the moment, because our first order flow formalism  accommodates also for black branes ($\kappa=0$, with $\sqrt{\eta}=dx dy$) and  black holes with hyperbolic horizon ($\kappa=-1$, $\sqrt{\eta} = \sinh \theta d \theta d \phi$). We integrate over a finite time interval, hence the factor $\beta_t$.
 
It turns out that the explicit form for the part in $\mathbf{\widetilde{g}}^2$ is:
 
 \begin{eqnarray}
S_2&=& \beta_t  \,\, \mathbf{\widetilde{ g}}^2\int d^3x \sqrt{\eta} \,\,\,
r^2e^{-K}
\left\{3\left[(re^{-K/2})\rq{}-\frac{\mathcal{W}}{\tilde \xi}\right]^2\right. \nn\\ 
&&\left.
-(r^4e^{-2K})\left(
\dot z^i+\frac{e^{K/2}}{r} g^{ik}\frac{\p_k \mathcal{W}}{\tilde \xi}
\right)g_{ij}\left(
\dot z^j+\frac{e^{K/2}}{r}g^{jl}\frac{\p_l \mathcal{W}}{\tilde \xi}\right)
\right\}+\nn\\
&& 
+ \,\, S^{(2)}_{td}\,,
\end{eqnarray}
 where the total derivative part is
\begin{equation}
S^{(2)}_{td} =  \beta_t  \,\,    \mathbf{\widetilde{ g}}^2   \int d^3x \sqrt{\eta} \ \left( -\frac34\frac{d^2}{dr^2}\left[
r^4e^{-2K}
\right]+ 2\frac{d}{dr}\left[(re^{-K/2})^{3} \frac{ \mathcal{W}}{\tilde \xi}\right]\right)\ .
\end{equation}

Also $S_0$ can be written as a sum of squares and total derivatives, plus a term whose variation vanishes once one enforces the first order equations. As done in \cite{Cardoso:2008gm}
we introduce harmonic functions $H_{\Lambda}$ of the form
\begin{equation}\label{accarm}
H_{\Lambda} = \tilde{a}_{\Lambda} + \frac{ \tilde{b}_{\Lambda}}{r}\,,
\end{equation}
and $S_0$ can be squared as:

\begin{eqnarray}
S_0
&=&  \beta_t\int d^3x \sqrt{\eta}\ \ \left\{{-}
2( \kappa\,\,r^2+c_1r)\left[M_{\Lambda}\,{}'-\frac{K'}{2}M_{\Lambda}-ie^{K/2}\frac{\tilde b_{\Lambda}}{r^2}\right]
\mc I^{\Lambda \Sigma}
\left[M_{\Sigma}\,{}'-\frac{K'}{2}M_{\Sigma}-ie^{K/2}\frac{\tilde b_{\Sigma}}{r^2}\right]+\right.\nn\\
&&\hspace{1cm}+2c_2\left[M_{\Lambda}\,{}'-\frac{K'}{2}M_{\Lambda}+\frac{M_{\Lambda}+ie^{K/2}\tilde a_{\Lambda}}{r}\right]
\mc I^{\Lambda \Sigma}
\left[M_{\Sigma}\,{}'-\frac{K'}{2}M_{\Sigma}+\frac{M_{\Sigma}+ie^{K/2}\tilde a_{\Sigma}}{r}\right]+\nn\\
&&\hspace{1cm}\left.+4 i c_1\frac{e^{K/2}}{r^2}\tilde b_{\Lambda}\mc I^{\Lambda \Sigma}\left[M_{\Sigma}+i\frac{e^{K/2}}2\left(\tilde a_{\Sigma}+\frac{\tilde b_{\Sigma}}r\right)\right] \right\} +S^{(0)}_{td}\,,
\end{eqnarray}
where $M_{\Lambda}$ is the lower part of the covariantly holomorphic vector $\mathcal{V}$ (further conventions and notation are in Appendix). For the electric solutions at hand the quantities $M_{\Lambda}$ are purely imaginary, hence the appearance of imaginary factors $i$ in the action.  The total derivative part is
\begin{eqnarray}\nonumber
S^{(0)}_{td} &=&  
        \beta_t \,\int  d^3x \sqrt\eta \ \ \bigg \{ \frac{d}{dr}\bigg[ -r^2  \left( \kappa +\frac{c_1}{r}+ \frac{c_2}{r^2} \right)  K' - c_1 K+ \\
        &+& 4 i\, e^{K/2} \left(  \big( \kappa+ \frac{c_1}{r} \big) \tilde b_{\Sigma} - \frac{c_2}{r} \tilde a_{\Sigma} \right) (\mc I^{-1})^{\Sigma \Lambda} M_ {\Lambda} - \frac{c_2}{r}{\bigg]}+ 2 q_{\Lambda} F^{\Lambda}_{tr} \bigg \}  \,. 
\end{eqnarray}

In performing the squaring we have made use of the special K\"ahler identities valid for purely real sections in Appendix \ref{AppA}, and of the following constraints between the charges $q_{\Lambda}$ and the parameters appearing in the harmonic functions \eqref{accarm}:
\begin{eqnarray}\label{VBHabel}
-\frac12 q_{\Lambda}(\mc I^{-1})^{\Lambda\Sigma}q_{\Sigma}=V_{BH}=-2 \left(\kappa\,\, \tilde b_{\Lambda}(\mc I^{-1})^{\Lambda\Sigma}\tilde b_{\Sigma} - c_1 \tilde a_{\Lambda} (\mc I^{-1})^{\Lambda\Sigma}\tilde b_{\Sigma}+ c_2  \tilde a_{\Lambda} (\mc I^{-1})^{\Lambda\Sigma}\tilde a_{\Sigma}\right)\ .
\end{eqnarray}
The last factor in the $S_0$ term is not a perfect square but it vanishes under variations with respect to $K$ and also under variations of $M_{\Lambda}$, provided that
\begin{equation}\label{formula}
M_{\Lambda} = -i e^{K/2} H_{\Lambda} 
\end{equation}
holds. 

At the end of the day, through the squaring of the action we found that a non-extremal electric solution in four dimensions satisfies the first order equations obtained by setting to zero each squared term in $S_0$ and $S_2$ :
\begin{equation}\label{zel}
{z^i}'  = - \frac{e^{K/2}}{ \tilde \xi  r} g^{ij} \partial_{j} \mathcal{W}\,, \qquad (r\,e^{-K/2})'  = \frac{\mathcal{W}}{\tilde \xi}\,,
\end{equation}
plus \eqref{formula} and \eqref{VBHabel}, for a superpotential $\mathcal{W}$ that satisfies \eqref{eqpot} with $V_g(z)$ given by \eqref{SKpotential}. Notice that these equations are analogous to those found by \cite{Cardoso:2008gm} in five dimensions.

Finally, we explicitly verified that the Einstein's equations and the scalars equations do not give further constraints. In other words, we verified that eq. \eqref{formula} \eqref{zel}, plus the form of the field strengths \eqref{ans_vec} are sufficient to solve all equations of motion. We provide the explicit proof of this fact in Appendix \ref{Einsteinapp} for the magnetic case - the electric case can be worked out in complete analogy.

\subsection{Magnetic configuration\label{magn-squaring}}

For the magnetic solution we start from the same ansatz for the warp factors \eqref{sol-param1}, that we repeat here for convenience:
$$
f(r)=\kappa +\frac{c_1}{r}+\frac{c_2}{r^2}+\mathbf{\widetilde{g}}^2  r^2e^{-2K(r)}\,.
$$
with $\mathbf{\widetilde{g}}$ as before.  The field strengths are magnetic
\begin{equation}\label{amagn2}
F_{tr}^{\Lambda}= 0 \,, \qquad
F_{\theta \phi}^{\Lambda} = \frac{p^{\Lambda}}{2} \sin \theta \,.
\end{equation}
Magnetic solutions found in \cite{Toldo:2012ec} can be cast in this form, as we will later show. Plugging this ansatz in the action  \eqref{action-CK}, we see that again we collect terms in $\mathbf{\widetilde{g}}^2$ and $\mathbf{\widetilde{g}}^0$:
\begin{equation}
S= S_0+S_2 
\end{equation}
and in this case we obtain:
\begin{equation}\nonumber
S_2 = \beta_t \, \mathbf{\widetilde{g}}^2 \int d^3x \sqrt{\eta} \, \bigg[ - \,   r^4 e^{-2K} \left( {z^i}' + \frac{e^{K/2}}{r} g^{ik} \frac{\partial_k \mathcal{W} }{\tilde \xi}\right) g_{ij} \left( {z^i}' + \frac{e^{K/2}}{r} g^{il} \frac{\partial_l \mathcal{W} }{\tilde \xi} \right)+
\end{equation}
\begin{equation}
	+3  r^2 e^{-K} \left( \frac{ \mathcal{W}}{{\tilde \xi}} -(re^{-K/2}) \right)^2 \bigg]+S^{(2)}_{td}\,,
\end{equation}
with 
\begin{eqnarray}
S^{(2)}_{td}= \beta_t \,  \mathbf{\widetilde{g}}^2 \int d^3x \sqrt{\eta} \,\left(  
-\frac34\frac{d^2}{dr^2}\left[
r^4e^{-2K}
\right]+\frac{d}{dr}\left[2(r^3e^{-3K/2})\frac{ \mathcal{W}}{{\tilde \xi}}\right]\right)\,.
\end{eqnarray}
Also in this case we introduce harmonic functions $$H^{\Lambda} = \tilde{a}^{\Lambda} + \frac{\tilde{b}^{\Lambda}}{r}\,,$$ so that the part $S_0$ can be squared as

\begin{eqnarray}
S_0 &=& \beta_t\int d^3x \sqrt{\eta} \ \ \left\{
2( \kappa\,\,r^2+c_1r)\left[L^{\Lambda}\,{}'-\frac{K'}{2}L^{\Lambda}+e^{K/2}\frac{\tilde b^{\Lambda}}{r^2}\right]
\mc I_{\Lambda \Sigma}
\left[L^{\Sigma}\,{}'-\frac{K'}{2}L^{\Sigma}+e^{K/2}\frac{\tilde b^{\Sigma}}{r^2}\right]+\right.\nn\\
&&\hspace{1cm}+2c_2\left[L^{\Lambda}\,{}'-\frac{K'}{2}L^{\Lambda}+\frac{L^{\Lambda}-e^{K/2} \tilde{a}^{\Lambda}}{r}\right]
\mc I_{\Lambda \Sigma}
\left[L^{\Sigma}\,{}'-\frac{K'}{2}L^{\Sigma}+\frac{L^{\Sigma}-e^{K/2} \tilde{a}^{\Sigma}}{r}\right]+\nn\\
&&\hspace{1cm}\left.+4c_1\frac{e^{K/2}}{r^2} \tilde{b}^{\Lambda}\mc I_{\Lambda \Sigma}\left[L^{\Sigma}-\frac{e^{K/2}}2\left( \tilde{a}^{\Sigma}+\frac{\tilde{b}^{\Sigma}}r\right)\right]\right\} + S^{(0)}_{td}\,,
\end{eqnarray}
with total derivative
\begin{eqnarray}\nonumber
S^{(0)}_{td} &=&  
        \beta_t \,\int  d^3x \sqrt\eta \ \ \bigg \{ \frac{d}{dr}\bigg[ -r^2  \left( \kappa +\frac{c_1}{r}+ \frac{c_2}{r^2} \right)  K' - c_1 K+ \\
        &+& 4 \, e^{K/2} \left(  \big( \kappa+ \frac{c_1}{r} \big) \tilde b^{\Sigma} - \frac{c_2}{r} \tilde a^{\Sigma} \right) \mc I_{\Sigma \Lambda} L^ {\Lambda} - \frac{c_2}{r} {\bigg]} \bigg \}  \,. \nonumber
\end{eqnarray}
Notice that, like in the electric case, one term is not a perfect square but its variation vanishes once the fields satisfy the first order equations.
In deriving the squaring we have made use once again of the identities of special geometry derived in Appendix \ref{AppA}
for real sections $L^{\Lambda}$. Furthermore, the charges need to satisfy the following constraint
\begin{eqnarray}\label{VBHab}
V_{BH}=-\frac12 p^{\Lambda}\mc I_{\Lambda\Sigma}p^{\Sigma} =-2\left( \kappa\,\, \tilde b^{\Lambda}I_{\Lambda \Sigma}\tilde b^{\Sigma}+
c_2 \tilde a^{\Lambda}I_{\Lambda \Sigma}\tilde{a}^{\Sigma}-c_1 \tilde b^{\Lambda}I_{\Lambda \Sigma} \tilde a^{\Sigma}
\right)
\end{eqnarray}
As in the electric case, there is a factor in the action that is not a perfect square, nonetheless it vanishes under variations with respect to $K$ and also $L^{\Lambda}$ provided that this holds:
\begin{equation}\label{defL}
L^{\Lambda} = e^{K/2}\left(\tilde a^{\Lambda}+\frac{ \tilde b^{\Lambda}}{r}\right) = e^{K/2} H^{\Lambda} \,.
\end{equation}
The first order equations coming from this squaring are given in the magnetic case by 
\begin{equation}\label{zmagn}
{z^i}'  = - \frac{e^{K/2}}{ \tilde \xi r} g^{ij} \partial_{j} \mathcal{W}\,, \qquad
(r\,e^{-K/2})'  = \frac{\mathcal{W}}{\tilde \xi} \,,
\end{equation}
with $\tilde{\xi} = 2 \sqrt{\xi_0 \xi_1^3} /3\sqrt3$ and $\mathcal{W}(z)$ satisfying \eqref{eqpot} with \eqref{SKpotential}. Also in this case the Einstein's and scalar equations of motion do not give further constraints, as shown in Appendix \ref{Einsteinapp}

\section{Non-extremal black holes in AdS}

In the previous section we have obtained a set of first order equations for asymptotically Anti de Sitter black holes, which fall in the class described by the metric ansatz
\begin{eqnarray}
ds^{2}&=&e^{K(r)}{f(r)}\,dt^2-{e^{-K(r)}}\left(\frac{dr^2}{f(r)}+ r^2 (d\theta^2+\sin\theta^2\,d\phi^2)\right)\ ,
\end{eqnarray}
with 
\begin{eqnarray}
f(r)&=&1+\frac{c_1}{r}+\frac{c_2}{r^2}+\tilde  g^{2}r^2e^{-2K(r)}\ .
\end{eqnarray}
As we noticed already, the only requirement for the squaring of the action and thus the derivation of the first order flow is that the covariantly holomorphic sections $L^{\Lambda}$ are purely real (and thus the symplectic dual sections $M_{\Lambda}$ are purely imaginary), so we can make use of the real special geometry relations in Appendix \ref{AppA}. This is true for theories with superpotentials of the form \eqref{sympl-frame} where the scalars are taken to be real.
Both purely electric and purely magnetic solutions, then, satisfy the first order flow for a \textit{real} scalar field defined as $z^1=\frac{X^1}{X^0}$, given by
\begin{eqnarray}\label{allflow}
\dot z^i&=&-(r e^{-K/2})^{-1} g^{ij}\frac{\p_j \mathcal{W}}{\tilde \xi}\ ,\qquad \left(r e^{-K/2}\right)\rq{} =\frac1{\tilde \xi}{\mathcal{W}}\ ,
\end{eqnarray}
in addition to an Hamiltonian constraint \eqref{VBHab} (or equivalently \eqref{VBHabel} for electric solutions).

\subsection{Black holes in the $t^3$ model}

From now on we focus on solutions of $\mc N=2$ Supergravity with Fayet-Iliopoulos electric gauging $g_{\Lambda}=\{g_0,g_1\}=g \xi_{\Lambda}$, with a single scalar parametrizing the nonlinear sigma model $SU(1,1)/U(1)$, described by the prepotential
$F=-2i\sqrt{X^0 (X^1)^3 }$. The solutions have zero axions i.e. real scalars, defined as $z=\frac{X^1}{X^0}$, and are expressed in terms of harmonic functions
\begin{eqnarray}\label{HH}
H_{\Lambda}=a_{\Lambda}+\frac{b_{\Lambda}}{r}\ ,\qquad \Lambda=0,1\ .
\end{eqnarray}
The solution for the warp factor $e^K$ is, both for electric and magnetic black holes,
\begin{eqnarray}
e^{-K(r)}=\beta^2\sqrt{H_0(H_1)^3}\ ,
\end{eqnarray}
where we explicitly introduced the dependence on an overall factor $\beta$, so that both the coefficients $a_{0}$ and $a_{1}$ are fixed by the solution. Indeed, notice that, since we are looking for solutions which asymptote to AdS, the solution of the radial flow has to be such that the scalar at infinity assumes the value that extremizes the gauging potential
\begin{eqnarray}
\p_z V_g|_\infty =0\qquad\rar\qquad  z_{\infty}= \frac{3\xi_0}{\xi_1}\ ,
\end{eqnarray} 
and the asymptotic cosmological constant is set by the value of the potential at infinity
\begin{eqnarray}
V_g(z_{\infty})=\Lambda=-\frac{3}{\ell_{AdS}^2}\ .
\end{eqnarray}
 In both the electric and magnetic case this requires that the parameter $\tilde g$ in the metric is
\begin{eqnarray}
 \tilde g=\frac{\hat{\xi}}{\beta}g\ ,\qquad 
\hat\xi=\frac{\sqrt2\xi_0^{1/4}\xi_1^{3/4}}{3^{3/4}}\quad \rar \quad \tilde \xi=\frac{\hat \xi}{\beta} \ ,
\end{eqnarray}
and thus the black hole solutions asymptote to an Anti de Sitter space with radius
\begin{eqnarray}\label{ellAdS}
\ell_{AdS}=\frac{1}{g\hat \xi }=\frac{1}{\beta \tilde g}\ .
\end{eqnarray}
The form of the metric is the same in both electrically charged and magnetically charged black holes; we proceed now give the scalar and gauge fields solutions for each configuration. The solutions describing non-extremal AdS black holes were respectively given in \cite{Duff:1999gh} and \cite{Klemm:2012yg,Toldo:2012ec}. 
In both cases the singularities are located at the zeroes of the function $e^{-K}$ while the horizons are at the zeroes of the function $f(r)$ (recall the form of the warp factors in \eqref{formuf}).

Finally let us mention that the first class, namely the electric configurations, are singular in the BPS limit \cite{Duff:1999gh}, while the BPS limit is regular for the magnetic ones, and correspond to a genuine extremal 1/4 BPS black hole configuration \cite{Cacciatori:2009iz,Dall'Agata:2010gj,Hristov:2010ri}.

\subsection{Electric solution \label{ElApp}}

The solution we present in this section is a reparameterization of the one found first in \cite{Duff:1999gh}.

The electrically charged black hole has scalar field
\begin{eqnarray}
z=\frac{X^1}{X^0}=\frac{H_0}{H_1}=\frac{a_0 r+b_0}{a_1 r+b_1}\ ,
\end{eqnarray}
and electric gauge fields
\begin{eqnarray}
A^{0}&=&\frac{\xi_1^{3/4}}{2\sqrt2 3^{3/4}\beta\xi_0^{3/4}}\frac{\sqrt{b_0(b_0-c_1a_0)+c_2a_0^2}}{a_0r+b_0} dt\ ,
\qquad
A^{1}=\frac{3^{1/4}\xi_0^{1/4}}{2\sqrt2 \xi_1^{1/4}\beta}\frac{\sqrt{b_1(b_1-c_1a_1)+c_2a_1^2}}{a_1r+b_1} dt\ ,\ \ \ \ \ \
\end{eqnarray}
or, in terms of electric charges
\begin{equation}\label{el-char}
q_0=\pm\frac{3^{3/4} \beta \sqrt{b_0(b_0-c_1a_0)+c_2a_0^2}}{\sqrt{2}}\ , \qquad 
q_1=\pm\frac{3^{3/4} \beta \sqrt{b_1(b_1-c_1a_1)+c_2a_1^2}}{\sqrt{2}}\ .
\end{equation}
The parameters $b_{\Lambda}$ are free while $a_{\Lambda}$'s are
\begin{equation}
a_\Lambda=\frac{\sqrt2}{3^{3/4}}\ell_{AdS}g_{\Lambda}\ .
\end{equation}
We can also invert the relation between the charges and the parameters as
\begin{eqnarray}
c_1&=&\frac{b_0}{a_0}+\frac{b_1}{a_1}+\frac{2}{3\sqrt3 \beta^2}\frac{a_0 a_1}{b_0 a_1-b_1 a_0}\left(\frac{q_1^2}{a_1^2}-\frac{q_0^2}{a_0^2}\right)
\ ,
\nn\\ 
c_2&=&\frac{b_0 b_1}{a_0 a_1}\left[
1+\frac{2}{3\sqrt3 \beta^2}\frac{a_0 a_1}{b_0 a_1-b_1 a_0}\left(\frac{q_1^2}{a_1b_1}-\frac{q_0^2}{a_0b_0}\right)
\right]
\,.
\end{eqnarray}
This solution satisfies the first order flow \eqref{zel} for a superpotential 
\begin{eqnarray}\label{Wel}
g \mathcal{W}=|g_{\Lambda}L^{\Lambda}|\ ,
\end{eqnarray}
with $L^{\Lambda}$ the symplectic sections that, in the case of real special geometry, are related to their symplectic duals by
\begin{eqnarray}
L^{\Lambda}=-i \mc I^{\Lambda \Sigma} M_{\Sigma}\ ,
\end{eqnarray}
which, as expected from the first order flow, can be written as 
\begin{eqnarray}
i\,M_{\Lambda}&=& e^{K/2} \left(
\tilde a_\Lambda+\frac{\tilde b_\Lambda}{r}
\right)\ ,
\end{eqnarray}
where the tilded parameters are related to our parametrization by
\begin{eqnarray}
\{\tilde a_\Lambda\,,\tilde b_{\Lambda}\}=\frac{3^{3/4}\beta}{2\sqrt2} \{a_{\Lambda}\,,b_{\Lambda}\}\ .
\end{eqnarray}
One can verify that for this $\tilde a_{\Lambda}$ and $\tilde b_{\Lambda}$ the charges \eqref{el-char} satisfy the constraint \eqref{VBHabel}.

Notice that for every Very Special geometry in $\mathcal{N}=2$ FI-gauged supergravity the quantity \eqref{Wel} is a superpotential, namely it satisfies eq. \eqref{eqpot}. Therefore we expect that a first order flow driven by this superpotential exists for zero axions solutions in every Very Special geometry with charges that satisfy the hamiltonian constraint \eqref{VBHab}. Turning on axions with a duality transformation will break the reality conditions on the sections we used in performing the squaring of the action, therefore it is not guaranteed that an analogous first order flow driven by \eqref{Wel} exists.

\subsection{Magnetic solution \label{MaApp}}

In analogy with what we did in the previous subsection for electric solutions, we present here the convenient reparameterization of the magnetic solution for the $t^3$ model found first in  \cite{Klemm:2012yg,Toldo:2012ec}.
The magnetic solution has a scalar field
\begin{eqnarray}
z=\frac{X^1}{X^0}=\frac{H_1}{H_0}=\frac{a_1 r+b_1}{a_0 r+b_0}\ ,
\end{eqnarray}
and magnetic gauge fields
\begin{eqnarray}
A^{\Lambda}&=&-\frac{1}{2}p^{\Lambda}\cos\theta d\phi\ ,
\end{eqnarray}
where the magnetic charges satisfy
\begin{eqnarray}\label{mag-char}
p^0&=& \pm \frac{\beta \sqrt{b_0(b_0-c_1a_0)+c_2a_0^2}}{\sqrt{2}}\ , \qquad
p^1=\pm\frac{\beta \sqrt{b_1(b_1-c_1a_1)+c_2a_1^2}}{\sqrt{2}}\ ,
\end{eqnarray}
$c_1$, $c_2$ are the real parameters entering the warp factor $f(r)$. The Einstein's equations are satisfied for coefficients $a_{\Lambda}$'s :
\begin{eqnarray}\label{a-magn}
a_0 = \frac{ \sqrt2 g \xi_1^{3/2} \ell_{AdS}}{3 \sqrt3 \sqrt{\xi_0}} =- \sqrt2 \mathcal{G}^0 \ell_{AdS}\,,  \qquad a_1 = \frac{\sqrt2 g \ell_{AdS} \sqrt{\xi_0 \xi_1}}{\sqrt3 } = -\sqrt2 \ell_{AdS} \mathcal{G}^1  \,,
\end{eqnarray}
that can be expressed in terms of "dual" gauging parameters  
\begin{eqnarray}
\mc G^{\Lambda}=(\mc I_{\infty}^{-1})^{\Lambda\Sigma}g_{\Sigma}\ ,\qquad 
\mc I_{\infty\,\Lambda\Sigma}\equiv \mc I_{\Lambda\Sigma}\big|_{z=z_{\infty}}\ .
\end{eqnarray} 
We can also choose to invert the relation between the physical charges and the coefficients $c_1$ and $c_2$ and obtain
\begin{eqnarray}
c_1&=&\frac{b_0}{a_0}+\frac{b_1}{a_1}+\frac{2}{ \beta^2}\frac{a_0 a_1}{b_0 a_1-b_1 a_0}\left(\frac{(p^1)^2}{a_1^2}-\frac{(p^0)^2}{a_0^2}\right)
\ ,
\nn\\ 
c_2&=&\frac{b_0 b_1}{a_0 a_1}\left[
1+\frac{2}{ \beta^2}\frac{a_0 a_1}{b_0 a_1-b_1 a_0}\left(\frac{(p^1)^2}{a_1b_1}-\frac{(p^0)^2}{a_0b_0}\right)
\right]
\,.
\end{eqnarray}
Such black hole solution, not only is a solution of the Einstein+Maxwell+Bianchi equations, but satisfies also the first order flow \eqref{zmagn} driven by a superpotential which  is NOT the supergravity one, $W_0(z)=g_{\Lambda}L^{\Lambda}$, but is given now by
\begin{eqnarray}\label{SuperW}
g \mathcal{W}=|\mc G^{\Lambda}M_{\Lambda}|=|L^{\Lambda}\mc I_{\Lambda\Sigma}\mc G^{\Sigma}|\ .
\end{eqnarray}
In the context of domain walls solutions, this function $\mc W$ is known as a \textit{fake superpotential}. For simplicity we will refer to both \eqref{Wel} and \eqref{SuperW} generically as superpotentials, defined by \eqref{eqpot} and by the first order flow \eqref{allflow}.

The sections $L^{\Lambda}$ can be written as 
\begin{eqnarray}
L^{\Lambda}&=& e^{K/2}{\beta} \left(
\tilde a_\Lambda+\frac{\tilde b_\Lambda}{r}
\right)\ ,
\end{eqnarray}
where the tilded parameters are related to our parametrization by
\begin{eqnarray}
\{\tilde a_\Lambda\,,\tilde b_{\Lambda}\}=\frac{\beta}{2\sqrt2} \{a_{\Lambda}\,,b_{\Lambda}\}\ .
\end{eqnarray}
One can verify that for this $\tilde a_{\Lambda}$ and $\tilde b_{\Lambda}$ the charges \eqref{mag-char} satisfy the constraint \eqref{VBHab}.

Notice that the existence of a superpotential of the form \eqref{SuperW} different than the one found in the elctric case \eqref{Wel} depends on the model taken into consideration and it is not guaranteed for any Very Special geometry.

\subsection{Duality relation between electric and magnetic solutions}

Let us discuss the action of a symplectic transformation on the theory.

Consider the matrix
\begin{eqnarray}\label{el-mag-transf}
\mc I_{\infty\,\Lambda \Sigma}&=&
\left(
\begin{array}{cc}
-z_{\infty}^{-3/2}&0\\0&-\frac{\sqrt{z_{\infty}}}{3}
\end{array}\right) \ .
\end{eqnarray}
Then, the symplectic transformation $S\in Sp(4,\mathbb R)$
\begin{eqnarray}
S=\left(
\begin{array}{cc}
0&-\mc I_{\infty}^{\Lambda \Sigma}\\ \mc I_{\infty\,\Lambda \Sigma}
\end{array}
\right)\ ,
\end{eqnarray}
generates a duality transformation on the symplectic sections
\begin{eqnarray}
\mc V\rar S\mc V\ ,
\end{eqnarray}
which corresponds to the reparametrization of the scalars
\begin{eqnarray}
z\rar\frac{z_\infty^2}{z}\ .
\end{eqnarray}

This transformation acts as a rotation from a purely electric to a purely magnetic frame. Indeed, consider the effective black hole potential appearing in the one dimensional Lagrangian for a purely magnetic configuration (see Appendix \ref{AppA} for the definition of black hole potential):
\begin{eqnarray}
V_{BH}&=& p^{\Lambda}\mc I_{\Lambda \Sigma}p^{\Sigma}=\mc Q^T_{mag}\mc M \mc Q_{mag}\ .
\end{eqnarray}
By the action of $S$ on the scalar sections, the matrix $\mc M$ transforms as $\mc M\rq{}=S^{T}\mc M\, S$, and the effective black hole potential becomes an electric effective potential
\begin{eqnarray}
V_{BH}&=& \mc Q^T_{mag}S^T\mc M\,S \mc Q_{mag}
=\hat q_{\Lambda}\mc I^{\Lambda \Sigma}\hat q_{\Sigma}\ ,
\end{eqnarray}
upon the identification
\begin{eqnarray}
\left(
\begin{array}{c}
0\\ \hat q_{\Lambda}
\end{array}
\right)&\equiv&
S\left(
\begin{array}{c}
p^{\Lambda}\\0
\end{array}
\right)=\left(
\begin{array}{c}
0\\0\\ -z_{\infty}^{3/2}p^0\\-\frac13z_{\infty}^{-1/2}p^1
\end{array}
\right)\ .
\end{eqnarray}
Thus, as known, the matrix $S$ rotates the magnetic to the electric configurations and the two are physically dual to each other. The same matrix $S$ provides the rotation to a magnetic frame if we start from purely electric charges $\mc Q_{el}=(0,q_{\Lambda})$.

Notice however that the gauging introduces a potential $V_g(z,\bar z)$, as defined in \eqref{Vgdef}\footnote{We focus now on the zero axions case, unless otherwise stated.}. The potential  for electric gauging is then
\begin{eqnarray}\label{Vg}
V_g(z,g_{\Lambda})&=&V_g(\frac{z_{\infty}^2}{z},g_{\Lambda})\ ,
\end{eqnarray}
which is \textit{invariant} under the duality action. However, it is still true that
\begin{eqnarray}
V_g(S\mc V,g_{\Lambda})&=&V_g(\mc V, S\mc G)\ ,
\end{eqnarray}
and in the zero axion, electric gauging case we have
\begin{eqnarray}
S\mc V=\left(\begin{array}{c}-\mc I_{\infty}^{\Lambda \Sigma}M_{\Sigma}\\ \mc I_{\infty\,\Lambda \Sigma}L^{\Sigma}\end{array}\right)\ ,\qquad S\left(\begin{array}{c}
0\\g_{\Lambda}
\end{array}
\right)=\left(\begin{array}{c}
-\mc I_{\infty}^{\Lambda \Sigma}g_{\Sigma}\\0\end{array}
\right)\ .
\end{eqnarray}
The gauging potential can be written in general for $U(1)$-gauged $\mc N=2$ Supergravity as 
\begin{eqnarray}
V_g=-3 |\mc L|^2+g^{i\bar\jmath}\p_i|\mc L|\p_{\bar\jmath}|\mc L|\ ,
\end{eqnarray}
where $\mc L=\langle\mc G,\mc V \rangle$. The duality transformation on the scalar sections acts on the potential by changing $\mc L$ to
\begin{eqnarray}
|\mc L|\rar |\mc L\rq{}|=\langle\mc G, S\mc V \rangle=\langle S\mc G, \mc V \rangle\ ,
\end{eqnarray}
which, starting from an electric gauging configuration gives
\begin{eqnarray}
 |\mc L\rq{}|=|\mc I_{\infty}^{\Lambda \Sigma}M_{\Sigma}g_{\Lambda}|\ .
\end{eqnarray}
This transformation leaves the gauging potential invariant and, in the case of zero axions, it generates a new superpotential $W\rq{}=|\mc L\rq{}|$ from the supersymmetric $W_0=|\mc L|$. Notice that if we interpret the gauging in the new theory as defined by the section $\mc L\rq{}$, the rotated theory has \textit{magnetic} gauging
\begin{eqnarray}
\hat g^{\Lambda}=-\mc I_{\infty}^{\Lambda \Sigma}g_{\Sigma}.
\end{eqnarray}

If one considers second order bosonic equations of motion, there is no difference between a magnetic (or electric) black hole solution in a purely electric-gauged theory, specified by ($\mc Q$, $g_\Lambda$), or again a magnetic (or electric) solution but now in a magnetic-gauged theory specified by ($\mc Q$, $\hat g^{\Lambda}$).
Put it differently, the duality-rotated solution does not care about the transformation of the section $\mc L$, and thus of the gauging. The potential \eqref{Vg}, indeed is, as stressed before, \textit{invariant} under the electric-magnetic duality matrix, and one can rotate a black hole solution to a dual one in the same gauged theory (i.e. the gauging is still given by the purely electric $g_{\Lambda}$\rq{}s). 

However, because the section $\mc L$ is the quantity defining the SUSY transformations of the fermionic fields, e.g. for the gravitino \cite{Andrianopoli:1996cm,Dall'Agata:2010gj}
\begin{equation}
\delta\psi_{\mu\,A}=D_{\mu}\epsilon_{A}+\varepsilon_{AB}\, T^{-}_{\mu\nu}\,\gamma^{\nu}\,\epsilon^{B}+\frac{i}{2}\,{\cal L}\, \delta_{AB}\,\gamma^{\nu}\,\eta_{\mu\nu}\,\epsilon^{B}\ ,
\end{equation}
as soon as one is interested in the BPS properties of the black hole extremal solutions, one has to specify which gauging is considered. 

In the electrically gauged theory the magnetic configuration is supersymmetric for a particular set of parameters that we discuss in the following subsection, while the electric solution does not have a supersymmetric limit. In light of the comments above, this is perfectly consistent with the electric-magnetic duality transformation since the dual solution of a BPS magnetic  black hole in an electric-gauged theory defined by ($p^{\Lambda}$, $\mc L$) is an electric black hole in a \textit{magnetic-gauged} ($\hat q_{\Lambda}$, $\mc L\rq{}$) theory, and not in an electric-gauged one.

The BPS equations are not invariant under the particular transformation \eqref{el-mag-transf} on the scalars. Among the duality transformations, however, there exist some that leave invariant the quantity $\mc L$ itself, and thus the supersymmetry equations. This has been studied in \cite{Halmagyi:2013uza} to generate black hole solutions with axions.

\subsection{The magnetic BPS black hole}

As stressed above, only the solution with magnetic charges admits an extremal BPS limit. This is achieved when the function $f(r)$ has a double pole or, more precisely, when it can be written as
\begin{eqnarray}\label{fextr}	
r^2f_{0}(r)&=&\frac{\beta^2}{\ell_{AdS^2}}(r^2-a^2)^2\ ,
\end{eqnarray}
with $r_h=a$ the horizon radius. This condition implies some restriction on the parameters of the metric $c_1$ and $c_2$, as we are going to explain
\footnote{
The choice of $c_2\neq0$ we have made throughout the paper allows us to get to the 
form of the warp factor as \eqref{fextr} at zero temperature. If the radial coordinate $r$ is chosen such that $c_2=0$, one would have to require that in the BPS case the warp factor has the  form $r^2f_{0}(r)=((r-r^*)^2-a^2)^2$, so that the coincident horizons would be $r_h=a+r^*$. }.

The supersymmetric solution satisfies one additional constraint on the parameters with respect to the non extremal one, and precisely
\begin{eqnarray}
g_{\Lambda}p^{\Lambda}=\pm1\ .
\end{eqnarray}
In particular, for the \lq\lq{}$-1$\rq\rq{} case, the parameters defining the BPS solutions are 
\begin{eqnarray}
c_1=\frac{8}{\ell_{AdS}^2}\left(\frac{b_1}{a_1}\right)^3\beta^2\ ,\qquad c_2= -3\left(\frac{b_1}{a_1}\right)^2+\frac{\ell_{AdS}^2}{4 \beta^2}+\frac{12}{\ell_{AdS}^2}\beta^2\left(\frac{b_1}{a_1}\right)^4\ ,
\end{eqnarray}
the horizon is
\begin{eqnarray}
r_h^2\equiv a^2=\frac{\ell_{AdS}}{2 \beta}\sqrt{1+4g_1p^1}\ ,
\end{eqnarray}
and we recall that the coefficient $a_1$ is fixed by \eqref{a-magn}.
One can choose to parametrize the BPS solution by $(p^1\ ,\ \xi_1\ ,\ \xi_0)$, or, equivalently, by  $(b_1\ ,\ \xi_1\ ,\ \xi_0)$, since the two parameters are related in the BPS limit by
\begin{eqnarray}
p^1=\frac{3}{4g_1}\left(
-1+\frac{4 \beta^2}{\ell_{AdS}^2}\frac{b_1^2}{a_1^2}
\right)\ .
\end{eqnarray}

Let us first notice an interesting fact. The 1/4-BPS solution satisfies the first order flow obtained in \cite{Dall'Agata:2010gj}, upon identification of the warp factors
\begin{eqnarray}
e^{\psi}=r\sqrt{f_0(r)}\ ,\qquad e^{U}=e^{K/2}\sqrt{f_0(r)}\ .
\end{eqnarray}
That was a gradient flow driven by the superpotential 
\begin{eqnarray}
\mc W=e^{U}|\mc Z-ie^{2(\psi-U)}\mc L|\ ,
\end{eqnarray}
however, since the BPS solution is just a particular case of the non-extremal set, it has to verify also a gradient flow driven by the magnetic superpotential $\mc W$ of eq. \eqref{SuperW}. 

We have explicitely verified that, on-shell, the magnetic superpotential and the BPS one are identical functions of $r$, as expected
\begin{eqnarray}
e^{-\psi(r)}\mc W(r)\equiv-\beta\ell_{AdS}W_{mag}(r)\ .
\end{eqnarray}
This raises questions about the nature of these BPS black holes, like possible relations to supersymmetric domain walls which are also solution of a first order flow driven by a superpotential satisfying eq. \eqref{eqpot}. 

We remark here another interesting characteristic of the BPS solutions that suggests they might be closely related to domain walls. Consider indeed the case of a magnetic black brane with an ansatz like the one of Sec. \ref{magn-squaring}, and whose BPS limit can be found in \cite{Toldo:2012ec}. In the black brane case the Supersymmetric constraint is simply
\begin{eqnarray}
p^{\Lambda}g_{\Lambda}=0\ ,
\end{eqnarray}
so the $p^{\Lambda}\rar0$ limit is well defined, and independent on $g_{\Lambda}$. The first order flow then reduces to
\begin{eqnarray}
U\rq{}(r)&=&e^{-U}\im(e^{-i \alpha}\mc L)\ ,\nn\\
\psi\rq{}(r)&=&2 e^{-U}\im(e^{-i \alpha} \mc L)\ ,\nn\\
\dot z^i&=& ie^{i \alpha}g^{i\bar\jmath} e^{-U}\bar D_{\bar\jmath}\bar{\mc L}\ ,
\end{eqnarray}
with phase $e^{i\alpha}=\pm ie^{i\alpha_{\mc L}}$,
thus yielding $\psi\rq{}=2U\rq{}$. Without loss of generality we can take then $\psi=2U$, which brings the metric ansatz to the form
\begin{eqnarray}
ds_{p=0}^2&=&e^{-2U}dr^2+e^{2U}\left(-dt^2+dx^2+dy^2\right)\ ,
\end{eqnarray}
which is the metric for a domain wall with BPS flow 
\begin{eqnarray}
U\rq{}(r)&=&\pm e^{-U}|\mc L|\ ,\nn\\
\dot z^i&=&\mp e^{i \alpha_{\mc L}}e^{-U}g^{i\bar\jmath}\bar D_{\bar\jmath}\bar{\mc L}\ ,
\end{eqnarray}
 governed by the superpotential $\mc W_{DM}=e^{2U}|\mc L|$ .

\subsection{Scaling symmetry}

The factor $\beta$ introduced in the parametrization above is not a physical parameter. Thanks to the scaling symmetry
\begin{eqnarray}
r\rar \lambda r \ , \qquad t\rar \frac{t}{\lambda}
\qquad c_1\rar \lambda c_1\ ,\qquad c_2\rar \lambda^2 c_2
\qquad b_{\Lambda}\rar \lambda b_{\Lambda}\ ,\qquad \beta\rar \frac{\beta}{\lambda} \ ,
\end{eqnarray}
one can set $\beta=1$ without affecting the solution. However, we find it convenient to present the solutions including $\beta$  since various parametrizations in the literature correspond to values of $\beta\neq1$. In particular, notice that one can easily go from dimensionful to dimensionless coordinates by choosing $\beta=1$ or $\beta=\ell_{AdS}$ respectively.

In order not to overload formulae with too many parameters, from now on we only discuss the case $\beta=1$.

\section{Holographic analysis of scalar field dynamics
 \label{Hol-Scalar}}

\subsection{The action for the canonical field}

We can choose a re-parametrization of the scalar field $z=e^{x \phi(r)+y}$, with $x$ and $y$ constant so that $\phi$ is a canonical normalized field. The constant $y$ allows to choose a reference value of the field in the $r$ flow. If we choose this to be the asymptotic infinity $r\rar+\infty$ we can study the fluctuations of $\phi$ with respect to the vacuum $AdS_4$, suitable for a holographic analysis. We will then use, in what follows
\begin{eqnarray}
\varphi(r)=\phi(r)-\phi(\infty)\ ,\qquad 
\phi(\infty)=\sqrt{3/8}\log[3\xi_0/\xi_1]\ .
\end{eqnarray}
In terms of this field the action becomes
\begin{eqnarray}\label{Action-normalized-scalar}
S&=&\int\sqrt{-g}d^4x\left(
\frac{R}2+\frac12\p_{\mu}\varphi(r)\p^\mu\varphi(r)-\left(\frac{3\xi_0}{\xi_1}\right)^{3/2}e^{\sqrt6\varphi}F^0_{\mu \nu}F^{0\,\mu \nu}+\right.\nn\\
&&\hspace{15pt}\left.
-3\left(\frac{3\xi_0}{\xi_1}\right)^{-1/2}e^{-\sqrt{2/3}\varphi}F^{1}_{\mu\nu}F^{1\,\mu \nu}-V(\phi)
\right)
\end{eqnarray}
with potential
\begin{eqnarray}
V(\varphi)=-\frac{3}{\ell_{AdS}^2}\textrm{Cosh}\left(\sqrt{\frac{2}{3}}\varphi\right)\ ,
\end{eqnarray}
and $\ell_{AdS}^{-2}=\sqrt{\frac{4}{27}\xi_0\xi_1^3}$. The field $\varphi$ is a massive scalar field with 
\begin{eqnarray}
m^2_{\varphi}=-\frac{4g^2\sqrt{\xi_0}\xi_1^{3/2}}{3\sqrt3}=-\frac{2}{\ell_{AdS}^2}\ ,
\end{eqnarray}

the dual operator conformal dimensions are $\Delta_{-}=1$, $\Delta_+=2$. The field satisfy the Breitenlohner-Friedman bound $m^2_{\varphi}{\ell_{AdS}^2}\geq-9/4$, moreover the mass is in the window $-9/4\leq m^2_{\varphi}{\ell_{AdS}^2}\leq -9/4+1$ which allows for Neumann and Mixed boundary conditions to be imposed at the asymptotic AdS \cite{Witten:2001ua}.

\subsection{Canonical radius}

The standard holographic analysis is carried out in coordinates for which the metric has an expansion $g_{tt}\sim \ell_{AdS}^{-2}(c+g^2r^2+\mc O(r^{-1}) )$, where $c$ is a constant. In our case this is achieved by shifting the $r$ coordinate and for our solutions this has the net effect of constraining the $b_{\Lambda}$ parameters to be $b_0=-3b_1 a_0 / a_1$.

\subsection{Asymptotic metric}
The metric ansatz we consider admits the asymptotic expansion in the canonical radius
\begin{eqnarray}
ds^2_{\infty}&=& \left(- d\tilde r^2+e^{2\tilde r/\ell}\left(
dt^2-\ell^2d\Omega_{(2)}^2\right) \right)\left(1+\mc O(e^{-2\tilde r/\ell}))
\right)
\end{eqnarray}
In the notations of \cite{Papadimitriou:2007sj} (see in particular eq. 3.1, 3.3), the metric asymptotes the AdS boundary as
\begin{eqnarray}\label{metricah}
ds^2\sim d\tilde r^2 +e^{2\tilde r/\ell}h_{(0)ij}(x)dx^idx^j
\end{eqnarray}
($\ell$ is the AdS radius, we dropped the suffix) and the field expansion in terms of the radial coordinate $\tilde r$ reads
\begin{eqnarray}\label{asympt-behavior-scalar}
\varphi\sim e^{-\Delta_-r/\ell}(\varphi_-(x)+...)+ e^{-\Delta_+r/\ell}(\varphi_+(x)+...)\ .
\end{eqnarray}
By comparison with the metric of our ansatz we find that $r$ and $\tilde r$ are related by $\frac{ r}\ell=e^{\tilde r/\ell}$, and 
\begin{eqnarray}
h_{(0)ij}(x)&=&\left(
\begin{array}{ccc}
1&0&0\\0&-\ell^2&0\\0&0&-\ell^2\sin\theta^2
\end{array}
\right)\ .
\end{eqnarray}

\subsection{Expansion of the scalar field}
The special geometry scalar field $z=X^1/X^0$ is related to the normalized real scalar as \begin{eqnarray}
z(r)=z_{\infty}e^{\sqrt{8/3}\,\varphi(r)}\ .
\end{eqnarray}
On the electric and magnetic solution this gives a radial profile 
\begin{eqnarray}
\varphi(r)=\epsilon\sqrt{\frac{3}{2}}\log\frac{H_1}{H_0}\ ,
\end{eqnarray}
with $\epsilon=1$ for the magnetic solution and $\epsilon=-1$ in the electric one.
The asymptotic expansion at infinity results, for a canonical radius with $b_0=-3b_1 a_0 / a_1$, in
\begin{eqnarray}
\varphi(r)\sim \frac{\alpha_1}{r}+\frac{\alpha_2}{r^2}+\mc O(r^{-3})\ ,\qquad as\ \ r\rar+\infty \ ,
\end{eqnarray}
$\alpha$ and $\beta$ give the value of expectation values and source for operators, depending on the choice of quantization.

For the electric and magnetic solutions at hand we can explicitely compute the values of $\alpha_1$ and $\alpha_2$ in function of the parameters appearing in the solution.
\begin{itemize} 
\item electric solution:
\begin{equation}
\alpha_1 = - \frac{\sqrt6\,\, b_1}{a_1} \,, \qquad \alpha_2 = - \frac{\alpha_1^2}{ \sqrt6}
\end{equation}
\item magnetic solution:
\begin{equation}
\alpha_1 =  \frac{\sqrt6\,\, b_1}{a_1}  \,, \qquad \alpha_2 = \frac{\alpha_1^2}{ \sqrt6}\,.
\end{equation}
\end{itemize}
If we introduce the parameter $\epsilon$ that takes the values $\epsilon=-1$ for electric, $\epsilon=1$ for magnetic solutions, we see that the boundary conditions are of the form
 \begin{equation}\label{form_Mult}
 \alpha_2 = \lambda \alpha_1^2\,, \qquad \lambda = \frac{\epsilon}{\sqrt6}\,.
 \end{equation}
 Boundary conditions of this kind are called {\it mixed boundary conditions} and in our particular case, they correspond in a triple trace deformation in the dual field theory \cite{Witten:2001ua,Papadimitriou:2007sj,Hertog:2004ns}, which falls in the class of ABJM models \cite{Aharony:2008ug}. In the dual field theory, the ABJM action $S_0$ is deformed by triple trace operators, 
\begin{equation}
S= S_0 + \lambda \int \mathcal{O}_1^3 \,,
\end{equation}
where $\mathcal{O}_1$ is an operator of conformal dimension one. An example of such operator $\mathcal{O}_1$ in 3 dimensions is a  bilinear of boundary scalars $\varphi$, transforming under the global R-symmetry group,
\begin{equation}
{\cal O}_1={\rm Tr}(\varphi^I a_{IJ}\varphi^J)\ ,
\end{equation}
for some constant matrix $a$  \cite{Hertog:2004ns,Hristov:2013sya}.

The holographic dictionary in presence of mixed boundary conditions has been worked out recently in \cite{Papadimitriou:2007sj} and it turns out that $\alpha_1$ is the vev of a dimension one operator in the dual field theory. The interpretation of the expectation value $\alpha_1$ as order parameter in the dual field theory allowed the  interpretation of the black hole phase transition of \cite{Hristov:2013sya} as a liquid-gas phase transition in the dual field theory.

By comparison with the asymptotic expansion \eqref{asympt-behavior-scalar}, we can then identify
\begin{eqnarray}
\varphi_-=\frac{\epsilon \sqrt6 Q_1}{\ell_{AdS}}\ ,
\qquad 
\varphi_+=\frac{\epsilon \sqrt6 Q_1^2}{\ell_{AdS}^2}\ , \qquad\qquad 
\varphi_+=\lambda \varphi_-^2\ ,
\end{eqnarray}
for the same \(\lambda\) as above and $Q_1=b_1/a_1$.

\subsection{First order flow for the normalized real scalar}
Both the electric and magnetic solutions satisfy a first order flow in terms of $\varphi$ given by
\begin{eqnarray}\label{eqfloww}
\varphi' &=&\frac{\ell_{AdS}}{r}e^{K/2}\p_{\varphi}W_{el,mag}(\varphi)\ ,\nn\\
(re^{-K/2})' &=&-\frac12\ell_{AdS} W_{el,mag}(\varphi)\ ,
\end{eqnarray}
where we recall that for the electric solution $\epsilon=-1$ and 
\begin{eqnarray}\label{WWel}
W_{el}(\varphi)=-\frac{2}{\ell_{AdS}}\left(
\frac34e^{\varphi/\sqrt6}+\frac14e^{-\sqrt{3/2}\varphi}
\right)\ ,
\end{eqnarray}
while for the magnetic solution $\epsilon=1$ and 
\begin{eqnarray}\label{WWmagn}
W_{mag}(\varphi)=-\frac{2}{\ell_{AdS}}\left(
\frac34e^{-\varphi/\sqrt6}+\frac14e^{\sqrt{3/2}\varphi}
\right)\ .
\end{eqnarray}
It is clear once again how the electric solution and the magnetic solution are related by 
\begin{eqnarray}\label{duality}
\varphi\rar-\varphi\ ,
\end{eqnarray}
The transformation \eqref{duality}, supplemented by the appropriate symplectic transformation acting on the electromagnetic charges (maintaining the Fayet-Iliopoulos parameter unaltered) leaves the potential and the one dimensional reduced action invariant and hence is a symmetry of the bosonic equations of motion. Such transformation on scalar field and charges transforms the electric solution into the magnetic one. 
However, the supersymmetry equations set further constraints on the charges, and these constraints are compatible with the presence of a horizon just in the magnetic case\footnote{If instead one wants to analyze the supersymmetry properties in a fully symplectic covariant setup, one should work in the framework of \cite{Dall'Agata:2010gj} and allow for duality transformation acting also on the Fayet-Iliopoulos parameters, as done in \cite{Halmagyi:2013uza}.}.

These superpotentials satisfy the relation
\begin{eqnarray}
V(\varphi)=\frac12\left(-\frac32 W(\varphi)^2+ (\partial_{\varphi} W(\varphi))^2\right)\ .
\end{eqnarray}
In order to determine the holographic properties of the solutions we have to expand the superpotential in terms of the field at $\varphi=0$. We obtain
\begin{eqnarray}\label{Whol-em}
W_{el,mag}(\varphi)&\sim&-\frac{2}{\ell_{AdS}}\left(
1+\frac{\varphi^2}{4}+\epsilon\frac{\varphi^3}{6\sqrt6}
\right)+\mc O(\varphi^4)\ .
\end{eqnarray}
Since the coefficient of the quadratic term is given by $- \Delta_{-} / (2 \ell_{AdS})$, the superpotential driving the flow belongs to the class of \lq{}\lq{}$W_-$\rq{}\rq{} superpotential (see Table 5 of \cite{Papadimitriou:2007sj}), {as expected since this is the class allowing for multi-trace deformations.}

\section{Holographic renormalization}

There exist nowadays well established procedures for computing the boundary counterterms and removing the divergencies, see for instance \cite{Bianchi:2001kw,deHaro:2000xn}. Here we will make use of the Hamilton-Jacobi (HJ) method, first used in the context of AdS/CFT by \cite{deBoer:1999xf}. The notion of energy and black hole mass in terms of the renormalized Brown-York \cite{Brown:1992br} boundary stress-energy tensor was analyzed first in \cite{Balasubramanian:1999re,Mann:1999pc,Emparan:1999pm} and the analysis suitable for black hole solutions such as ours is the one of Papadimitriou \cite{Papadimitriou:2007sj}, where the presence of mixed boundary conditions was taken into account in the renormalization of the stress-energy tensor. We follow closely this procedure. Let us finally mention that the analysis of the mass obtained from the HJ renormalization technique was performed in
 \cite{Batrachenko:2004fd}, in which the authors compute the mass for electric black holes solutions in the truncation of $\mc N=8$ $SO(8)$-gauged theory to the $\mc N=2$ $U(1)$ gauged subsector.

\subsection{Regularized action}

To properly compute the conserved quantities in a 4-dimensional spacetime with boundary $\p\mc M$ we have to consider the bulk action together with the contribution coming from the Gibbons-Hawking boundary term
\begin{eqnarray}\label{IplusGH}
I&=&I_{bulk}+I_{GH}=\nn\\
&=& \int_{\mc M}  d^4x \sqrt{-g}\ \left(\frac R2+g_{i j}\p_{\mu}z^i\p^{\mu} z^{j}+ I_{\Lambda \Sigma} F_{\mu \nu}^{\Lambda}F^{\mu \nu\,\Sigma}-V_g\right) - \int_{\p\mc M}d^3 x\sqrt{h}\Theta\,.
\end{eqnarray}
In the Gibbons-Hawking term $\Theta$ is the trace of the extrinsic curvature
\begin{eqnarray}
\Theta_{\mu \nu}=-\frac{1}{2}(\nabla_\mu n_{\nu}+\nabla_\nu n_{\mu})\equiv - \nabla_{(\mu}n_{\nu)}
\end{eqnarray} 
where we choose $n^{\mu}=(0,\sqrt{-g^{rr}},0,0)$ as an outward-pointing normal vector to $\p \mc M$, and $h=\det(h_{\mu \nu})$ is the determinant of the induced metric $h_{\mu \nu}=g_{\mu \nu}+n_{\mu}n_{\nu}$ on $\p \mc M$ \cite{Brown:1992br}.

For any Killing vector field $K^{a}$ associated with an isometry of the boundary induced metric $h_{\mu \nu}$, we can define the conserved quantity
\begin{equation}\label{Qcons}
Q_{K} = \frac{1}{8 \pi} \int_{\Sigma} d^2x \sqrt{\sigma} u_{a} \tau^{ab} K_{b}\,,
\end{equation}
where $\Sigma$ is the spacelike section of the boundary surface $\p\mc M$, $u^{a}=\sqrt{h^{tt}}(1,0,0)$ is the unit normal vector to $\Sigma$ in $\p\mc M$, $\sigma_{ab}$ is the induced metric on $\Sigma$ and finally the local surface energy momentum tensor is defined as the variation of the boundary action with respect to the induced metric
\begin{equation}
\tau^{ab} = \frac{2}{\sqrt{h} } \frac{ \delta I}{ \delta h_{ab}} \ .
\end{equation}
The mass of the black hole is the conserved quantity associated with the Killing vector $K^{a}=(1,0,0,0)$ of the metric $h_{\mu \nu}$ at the boundary. 

Notice that, since the boundary stress energy tensor computed for the action \eqref{IplusGH} is divergent, we need to regulate it and then add an appropriate counterterm action $I_{ct}$:
\begin{eqnarray}
I=I_{reg}+I_{ct}\ ,
\end{eqnarray}
or equivalently
\begin{eqnarray}
\tau^{ab}=\tau^{ab}_{reg}+\tau^{ab}_{ct}\ .
\end{eqnarray}

We choose to regularize \eqref{IplusGH} by introducing a cutoff radius $r_0$ in the parametrization of the spacetime, thus leaving a truncated spacetime $\mc M_0$ with boundary $\p\mc M_0$ located at $r=r_0$. Removing the cutoff corresponds to taking the limit $r_0\rightarrow\infty$.
The regulated boundary stress tensor receives contribution from the Gibbons-Hawking term and has the form
\begin{eqnarray}
\tau^{ab}_{reg}=\frac{2}{\sqrt{h} } \frac{ \delta I}{ \delta h_{ab}}\Big|_{r_0} =  \left( \Theta^{ab}-\Theta h^{ab} \right)\Big|_{r_0}\ .
\end{eqnarray}
The mass of the black hole solution is the finite on-shell quantity remaining after removing the cutoff in the expression
\begin{eqnarray}\label{Mren}
M_{ren}=Q_{K}(\tau^{ab}_{reg})+Q_{K}(\tau^{ab}_{ct})\equiv E_{reg}+E_{ct}\ .
\end{eqnarray}
We will discuss in the rest of the section how to compute the contribution from the counterterms and how to derive a finite formula for the black hole mass.

\subsection{Canonical counterterms}

As said before, we have to renormalize the boundary stress energy tensor in order to extract finite quantities like the mass.
The counterterms needed to subtract the divergences come from an action of the form
\begin{eqnarray}\label{Ict}
I_{ct,can}=\int_{\p\mc M_0}d^3x\sqrt{h}\, \left(
- W(\varphi)+Z(\varphi)\mc R
\right)
\end{eqnarray}
where $\cal{R}$ is the Ricci scalar of the 3-dimensional boundary metric $h_{ij}$, $W(z)$ satisfies the relation
\begin{eqnarray}
V(\varphi)=\frac12\left(-\frac32 W(\varphi)^2+(\partial_{\varphi}W(\varphi))^2\right)\ ,
\end{eqnarray}
and determine $Z(\varphi)$ as from the equations (6.3)-(6.6) of \cite{Papadimitriou:2007sj}. 

The general solution for $W(\varphi)$ has been derived in \cite{Papadimitriou:2007sj} for a scalar potential like the one we are considering in \eqref{Action-normalized-scalar} and reads:
\begin{equation}\label{WWnu}
W_{\nu}(\phi)= - \frac{2}{\ell_{AdS}} \frac{1}{(1-\rho^2)^{3/4}} \frac{1-\rho^2+\sqrt{1+2 \nu \rho +\rho^2}}{\sqrt{2(1+ \nu \rho + \sqrt{1+ 2 \nu \rho +\rho^2})}}\,,
\end{equation}
where
\begin{equation}
\rho= \tanh \left(\sqrt{\frac23} \varphi \right) \qquad {\text and} \qquad \nu \geq -1\,.
\end{equation}
In other words, there is one parameter family of solutions for $W$ depending on an arbitrary real parameter $\nu$. 
The general superpotential $W_{\nu}(\varphi)$ admits the following series expansion in terms of the scalar field at infinity
\begin{eqnarray}\label{W-nu-exp}
W_{\nu}(\varphi)&=&-\frac{2}{\ell}\left(
1+\frac{\varphi^2}{4}+\frac{\nu}{6\sqrt6}\varphi^3+\mc O(\varphi^4)
\right)\ ,\qquad\qquad \textrm{as}\ \ r\rar\infty\ .
\end{eqnarray}
For any finite value of $\nu \geq -1$, the coefficient of the quadratic term in $\varphi$ is $ - \Delta_{-}/(2 \ell_{AdS})$. Therefore, the function $W$ is of the type ''$W_{-}$'', in the conventions of \cite{Papadimitriou:2007sj} and it is suitable for removing the divergencies from the action.

The first order flow derived in Sec. \ref{sec-squaring} is driven by superpotentials in the class of \eqref{WWnu}. In particular, the superpotential of the electric solution \ref{WWel} corresponds to the choice  $\nu_{el}=-1$ in \ref{WWnu}, and the magnetic one \ref{WWmagn} to the choice  $\nu_{mag}=1$.

Moreover, from \eqref{W-nu-exp} one can see that the term of order $\varphi^3$ gives a finite contribution that depends on the parameter $\nu$. Following the procedure of \cite{Papadimitriou:2007sj}, we are going to include in the canonical counterterms only the divergent terms. Every finite contribution is considered separately, and will be discussed in the following subsection.

The counterterms \eqref{Ict} are responsible for the renormalization of the boundary stress tensor. 
They give the contribution
\begin{eqnarray}\label{conserved-charge}
 Q_{\xi}(\tau^{ab}_{ct,can})\equiv E_{ct,can}= \frac12 \sqrt{h}h_{tt}\left[
h^{tt}W_{\nu=0}+
Z\left(\mc R\,h^{tt}+2\mc R^{tt}
\right)\right]\ ,
\end{eqnarray}
whose explicit expression depends on the solution $\varphi(r)$, which is a priori different for the electric and magnetic case. 
However, the divergent part of the counterterms is universal for a potential of the form \eqref{eqpot}: 
\begin{eqnarray}\label{Ect}
E_{ct,can}^{e,m}= \left[ \ell^{-2}r^3+(1-3Q_1^2  \ell^{-2})r
+\left(\frac{c_1}{2}-3Q_1^3 \ell^{-2}\right)
\right]+ \mc O(r^{-1})\ .
\end{eqnarray}
with $Q_1 = b_1 \xi_0^{1/4}/ \xi_1^{1/4} = b_1/a_1$ for the electric solution, and $Q_1 = b_1 \xi_1^{1/4}/ ( 3^{1/4} \xi_0^{1/4}) = b_1/a_1$ for the magnetic one.

\subsection{Finite terms}

Mixed boundary conditions for the solutions at hand correspond to a multi-trace deformation of the dual field theory. In order for the holographic renormalization procedure to have a well-defined variational principle, finite terms $I_{fin}$ have to be added to the action, accordingly to the prescription of \cite{Papadimitriou:2007sj}
\begin{equation}\label{addition-fin}
I = I_{reg} + I_{ct,can} + I_{fin}\,,
\end{equation}
where the finite part $I_{fin}$ in \eqref{addition-fin} is defined as
\begin{equation}
I_{ct,fin} = \int_{\p\mc M_0}d^3x\sqrt{h_{(0)}}\,  \,\tilde f(\varphi_-)\ ,
\end{equation}
where $h_{(0),ij}$  is defined as in eq. \eqref{metricah}. For a scalar field with mass and asymptotic expansion as in Sec. \ref{Hol-Scalar} this takes the form 
\begin{eqnarray}\label{fin-ct}
\tilde f'(\varphi_-)=-\hat\pi_{\Delta_+}(\varphi_-)=\frac{1}{\ell}\lambda\varphi_-^2\ ,\qquad \tilde f(\varphi_-)= \frac{ \lambda}{3\ell}\varphi_-^3=2\frac{Q_1^3}{\ell^4}\ .
\end{eqnarray}
Since $\tilde f(\varphi_-)\propto \varphi_-^{d/\Delta_-}$ ($d$ is the dimension of the boundary, in our case $d=3$), the mixed boundary conditions of the black holes solutions in this paper lead to a conformal dual theory and describe a marginal multi-trace deformation.

The finite counterterm \eqref{fin-ct} is responsible for a shift in the regularized stress energy tensor given by
\begin{eqnarray}\label{formlast}
E_{fin}= \frac{1}{8 \pi} \int_{\p\mc M_0}\sqrt{h_{(0)}}\, h^{(0)}_{tt} \,\tilde f(\varphi_-)\ ,
\end{eqnarray}
and computed on our solutions it yields
\begin{eqnarray}\label{Efin}
E_{fin}= \frac{1}{8 \pi} \int_{\p\mc M_0}\sqrt{h_{(0)}}\, h^{(0)}_{tt} \, \tilde f(\varphi_-)= \frac12 \frac{\sqrt{h_{(0)}}}{\sin\theta}h_{(0)tt}\tilde f(\varphi_-)= \ell_{AdS}^{-2}Q_1^3\ .
\end{eqnarray}
Notice there is no dependence on the parameter $\nu$ from the finite term of $W(\varphi,\nu)$ \eqref{W-nu-exp}. Indeed, the prescription for the finite terms that give a well-defined boundary problem makes sure this ambiguity is eliminated and no $\nu$-dependence appears in the total finite term \eqref{Efin}.
It is important to notice that this same contribution is precisely the finite term coming from the superpotential \eqref{W-nu-exp} when $\nu$ is chosen according to the solution, that is when one chooses as the counterterm $W$ exactly the superpotential that drives the first order flow of the non-extremal solution. 

Two more comments are in order. 1) In our case, the finite term given by holographic renormalization coincides with the finite term of the counterterm superpotential, when chosen as $W_{cterm}\equiv\mc W_{flow}$. We are now going to motivate this statement. We expand for $r \rightarrow \infty$  the right hand side and the left hand side of equation
\begin{equation}
\varphi' =\frac{\ell}{r}e^{K/2}\p_{\varphi}W_{\nu}(\varphi)\,
\end{equation}
and we get:
\begin{eqnarray}
\frac{\alpha_1}{r^2} +\frac{ 2 \alpha_2}{r^3} + O(r^{-3}) &= & \frac1r \left(\varphi + \frac{\nu \varphi^2}{ \sqrt6 } + O(\varphi^3) \right)   \,.
\end{eqnarray}
Expanding further the right hand side and using \eqref{form_Mult}, we obtain a relation between $\lambda$ and $\nu$:
\begin{equation}
\alpha_2 = \lambda \alpha_1^2 = \frac{1}{\sqrt6} \nu \alpha_1^2 \qquad \rightarrow \qquad  \lambda = \frac{1}{ \sqrt6} \nu \,.
\end{equation}
In other words, the boundary conditions of the scalar field, namely the function $f$, is related to the parameter $\nu$ appearing superpotential $\mathcal{W}_{\nu}$ generating the flow. Now, since $\varphi_{-} = \alpha_1/\ell$, we have $\tilde{f} =\frac {\lambda \alpha_1^3}{3 \, \ell^4}$ so that, by means of formulas \eqref{formlast} and \eqref{Ict}
\begin{equation}
E_{fin, f} = \frac{ \nu \alpha_1^3}{ \ell^2 6 \sqrt6}\,, \qquad E_{fin,\mathcal{W}} = \frac{\nu \alpha_1^3} {\ell^2\, 6 \sqrt6}\,.
\end{equation}
We see that the finite contributions to the mass exactly coincide. The fact that the finite term from holographic renormalization coincides with the finite term of the counterterm superpotential when $W_{cterm}\equiv\mc W_{flow}$ is a property that holds for any marginal multi trace deformation\footnote{We thank I. Papadimitriou for clarifying correspondence on this point.}.

2) This explains the choice of counterterms of e.g. \cite{Batrachenko:2004fd} in the computation of the mass of the electric black hole. However, that choice is not universal for all black hole solutions of the same theory \eqref{action-CK}: thanks to the derivation of the first order flow for the magnetic solution and its corresponding superpotential in Sec. \ref{sec-squaring}, we have now an example where $\mc W_{mag}\neq \mc W_{el}$, thus the counterterm should be chosen differently. Without the knowledge of the first order flow (thus of the superpotential) one should proceed with the computation of the finite terms as done above in this section, following \cite{Papadimitriou:2007sj}.

\subsection{Black hole mass}

In order to compute the energy we need first of all to compute, in the same way, the contribution from the Brown-York boundary term to the energy, regularized by a cutoff $r_0$. By using the same definition \eqref{conserved-charge}, taking into account also the finite terms \eqref{Efin}, we obtain
\begin{eqnarray}\label{Ereg}
E_{reg}+E_{fin}=\frac{1}{\ell^2}\left[-r^3+(3Q_1^2 -\ell^2)r+(3Q_1^3 -\ell^2 c_1)\right]+\mc O(r^{-1})
\end{eqnarray}

From \eqref{Ect} we can now subtract the divergences from the regularized energy \eqref{Ereg} so that the renormalized mass, obtained from a well defined variation principle, is
\begin{eqnarray}\label{Mass}
M_{el,mag}=E_{reg}+E_{ct}+E_{fin}=
-\frac{c_1}{2}\ .
\end{eqnarray}
We have verified that the expression \eqref{Mass} obtained for the black hole mass satisfies the standard first law of thermodynamics
\begin{equation}\label{first_law}
	{\rm d} M = T {\rm d} S  +\phi^{\Lambda}  {\rm d} q_{\Lambda} - \chi_{\Lambda} {\rm d} p^{\Lambda}\ ,
\end{equation}
where $\phi^{\Lambda}$ and $\chi^{\Lambda}$ are respectively the electrostatic and magnetostatic potentials. The expression for the mass used in the thermodynamics analysis of \cite{Hristov:2013sya} coincides with the value of \eqref{Mass}. Moreover, this expression coincides with the value obtained via the Ashtekar-Magnon-Das prescription, as shown in Appendix \ref{AppADM}.

\subsection{Renormalized on-shell action}

\subsubsection{Magnetic solution}

We consider here the magnetic black hole solution, whose mass is given by eq. \eqref{Mass}. From the results of the previous Section \ref{Hol-Scalar}, the thermodynamical potential is\footnote{Here and in the following it should be understood that $Q_0=-3Q_1$} 
\begin{eqnarray}\label{Omega} 
\Omega&=& \frac{\Gamma}{\beta_t} =\frac{I_{reg}}{\beta_t}+ E_{ct}+E_{fin}=
	-\frac{3 c_1}{4}-\frac{r_+}{2}+\frac{1}{\ell^2}(2 Q_1-r_+)(Q_1+ r_+)^2
\end{eqnarray}
where $I_{reg}$ is the regularized on shell action, $I_{ct}=\beta_t E_{ct}$ is the counterterm action, and $r_+$ is the radius of the outer horizon, defined by the relation
	\begin{eqnarray} \label{radiuseventhorizon}
	c_2 = -(c_1+r_+)r_+ - \frac1{\ell^2}(r_+-3Q_1) (Q_1 + r_+)^3 \ .
	\end{eqnarray}
We want to find the thermodynamic relation satisfied by the renormalized on-shell action, interpreted as the free energy of the system, giving information on the thermodynamic ensemble corresponding to the black hole solution.
The relevant thermodynamic quantities are
\begin{eqnarray}
T&=&\frac{1}{4\pi}e^{K}\left.\frac{df(r)}{dr}\right|_{r_+}\ ,\qquad 
S=\pi r_{+}^2 e^{-K}\Big|_{r_+} \ ,
\end{eqnarray}
where $f(r)$ is defined in \eqref{sol-param1} so that 
\begin{eqnarray}\label{TS}
TS=  \frac{c_1}4 +\frac{r_+}{2}-\frac1{\ell^2} (2 Q_1 -  r_+)(Q_1 + r_+)^2 
\ ,
\end{eqnarray}
and the mass is, as given in eq. \eqref{Mass},
\begin{eqnarray}
M=-\frac{c_1}2\ .
\end{eqnarray}
The explicit expressions \eqref{TS}, \eqref{Mass} allow to verify the identity 
\begin{eqnarray}
\Omega_{mag}=M-TS\, ,
\end{eqnarray}
together with the first law \eqref{first_law} for the magnetic solutions
\begin{equation} \label{thermo-law}
{\rm d }M=T {\rm d }S- \chi_\Lambda {\rm d }p^{\Lambda}  \ .
\end{equation}
The magnetostatic potentials $\chi_\Lambda$  are defined as the value of the dual electric fields at infinity
$A_{\Lambda\, t}(r)$, in a gauge where the gauge field vanishes at the horizon \cite{Hawking:1995ap}. They can be computed from the dual field strengths $G_\Lambda$
\begin{equation}
G_{\Lambda\, rt}=\frac{e^{K(r)}}{2r^2}\mc I_{\Lambda \Sigma}(z(r))p^{\Sigma}\ ,
\end{equation}
with $\partial_r A_{\Lambda\, t}(r)= G_{\Lambda\, rt}(r)$. One finds
\begin{eqnarray}
\chi_{\Lambda}=\left(g_0^2 \ell^2
\frac{p^0}{(-3Q_1+r_+)}\ ,\ g_1^2\ell^2\frac{p^1}{3(Q_1+r_+)}
\right)\ ,\end{eqnarray}
where magnetic charges have been given in \eqref{mag-char} and can be rewritten as\footnote{Gauging parameters $g_\Lambda$ and $\mc G^\Lambda$ are related as
\begin{align}
\mc G^0&= \frac{-1}{2g_0\ell^2}\ , \qquad \mc G^1=\frac{-3}{2 g_1 \ell^2}\ .
\end{align}}
\begin{eqnarray}
p^0&=& \pm  \ell \mc G^0  \sqrt{c_2-3Q_1(-3Q_1-c_1)}\ ,\nn\\
p^1&=& \pm \ell \mc G^1 \sqrt{c_2+Q_1(Q_1-c_1)}\ .
\end{eqnarray}
The combined form of the thermodynamic potential \eqref{Omega}, together with the first law \eqref{thermo-law}, yields
\begin{eqnarray}
{\rm d } \Omega= -S {\rm d }T - \chi_\Lambda {\rm d }p^{\Lambda} \ .
\end{eqnarray}
This relation is stating that the magnetic black hole is a state in a canonical thermodynamic ensemble. Indeed, given our choice of boundary terms and renormalization scheme, the variation of the action yields the equations of motion when \emph{magnetic charges} are kept fixed. One interprets the thermodynamic potential $\Omega$ as a function of  temperature $T$ and  magnetic charges $p^{\Lambda}$, $\Omega_{mag} = \Omega_{mag}(T, p^{\Lambda})$. This is different from purely electric black holes, that we now turn to analyze.
\vspace{3mm}

\subsubsection{Electric solution}
Let us review the free energy computation for purely electric black holes. In this case, the renormalized on-shell action takes the form 
\begin{eqnarray}\label{Omega_el}
\Omega&=&\frac{\Gamma}{\beta_t} =\frac{I_{reg}}{\beta_t}+ E_{ct}+E_{fin}=\frac12 \left(\frac{c_1}{2} +r_+ \right)
	\end{eqnarray}
For a choice of gauge such that $A_t^{\Lambda}(r_+)=0$, the electric chemical potentials are defined as the value of the gauge field at infinity \cite{Hawking:1995ap}
\begin{eqnarray} 
\phi^{\Lambda}=\left(
 \frac{1}{4 g_0^2 \ell^2} \frac{q_0}{(r_+ -3 Q_1)}\,,
\frac{3}{4 g_1^2 \ell^2} \frac{q_1}{(r_+ + Q_1)}
\right)\ .
\end{eqnarray} 
Notice that the coefficients $c_1$ and $c_2$ in $f(r)$ are related to the charges by
\begin{eqnarray}
q_0&=& \pm \ell g_0  \sqrt{c_2-3Q_1(-3Q_1-c_1)}\ ,\nn\\
q_1&=&  \pm \ell g_1 \sqrt{c_2+Q_1(Q_1-c_1)}\ ,
\end{eqnarray}
and to the horizon radius by \eqref{radiuseventhorizon}, giving
\begin{eqnarray}\label{chip}
\phi^\Lambda q_\Lambda=
-c_1-r_++\frac1{\ell^2} (2 Q_1 -  r_+)(Q_1 + r_+)^2 
\ .
\end{eqnarray}
Hence, by making use of  \eqref{TS}, \eqref{Mass} and \eqref{chip}, we have that 
\begin{eqnarray}
M-TS- \phi^{\Lambda} q_{\Lambda} &= & -\frac{c_1}{2} - \left(\frac{c_1}4 +\frac{r_+}{2}-\frac1{\ell^2} (2 Q_1 -  r_+)(Q_1 + r_+)^2 
\right) \nonumber \\
&+ & c_1 + r_+-\frac1{\ell^2} (2 Q_1 -  r_+)(Q_1 + r_+)^2 = \frac{c_1}{4} +\frac{r_+}{2}\ ,
\end{eqnarray}
so
\eqref{Omega_el} can be recast as
\begin{eqnarray}
\Omega_{el}=M-TS- \phi^{\Lambda} q_{\Lambda}
\end{eqnarray}
which is the statement that the renormalized on shell action coincides with the Gibbs Free energy, $\Omega_{el} = \Omega_{el} (T, \phi^{\Lambda})$. The variation of the action, given the counterterms we added, gives the equations of motion only if the \emph{electric chemical potentials} are kept fixed, hence the chosen ensemble is grand-canonical.
We have checked explicitely that the first law for electric configurations is satisfied
\begin{equation}
{\rm d }M = T {\rm d }S + \phi^{\Lambda} {\rm d }q_{\Lambda}\,,
\end{equation}
and the thermodynamic potential is extremized at fixed temperature and chemical potentials:
\begin{equation}
{\rm d } \Omega_{el} = -S {\rm d } T - q_{\Lambda} {\rm d } \phi^{\Lambda}\,.
\end{equation}

Let us conclude with a remark. For an electric configuration, it is possible to change thermodynamic ensemble to the canonical one, i.e. where charges $q_\Lambda$ are kept fixed instead of chemical potentials, by adding the Hawking-Ross boundary term\cite{Hawking:1995ap}, which reads:
\begin{eqnarray}
I_{HR}=  - \int_{\p\mc M_0} d^3x \, \sqrt{h} \, F^{\mu \nu} n_{\mu} A_{\nu}\,.
\end{eqnarray}
This is equivalent to performing a Legendre transform on $\phi^{\Lambda}$. The resulting free energy obtained upon addition of the Hawking-Ross counterterm is, as expected, $\Omega = M- TS$.

\section*{Conclusions and outlook}

The work in this paper is an investigation on black holes solutions in FI gauged supergravity. 

By exploiting a squaring of the action \lq\lq{}\`a la BPS\rq\rq{} we presented a first order formulation of electric and magnetic black holes coupled to a real scalar field in a Supergravity potential, and we have identified the superpotential for each configuration. 

Electric and magnetic black holes have been discussed with a symplectic covariant formalism which allows to understand that the duality rotation is still consistent on the non-extremal solutions but does not preserve the Supersymmetry properties of the extremal one. The supersymmetric solutions, moreover, have been revealed to satisfy also a first order flow when the superpotential is the same of the supersymmetric domain walls, giving new insights on the nature of these solutions.

The mass of the black hole has been computed for both electric and magnetic black holes through the techniques of holographic renormalization in presence of mixed boundary conditions for the scalar fields. In particular, it has been stressed that there is no ambiguity in the finite terms that the renormalization procedure requires, more precisely they are determined uniquely by the superpotential driving the first order flow. 
The mass formula obtained obeys the first order law of thermodynamics and the
thermodynamics relation \eqref{thermo-law} between the potential and the mass is satisfied.

The first order, superpotential formulation of the solutions could be useful for constructing new black holes of U(1) gauged Supergravity. The same formulation could possibly be generalized to the case with axions upon a suitable complexification of the equations, or to understand the string/M- theory origin of these black holes \cite{Cvetic:1999xp,Halmagyi:2013sla}.

Finally, it is known that the first order formalism of  fake supergravity for domain walls solutions (see for example \cite{Freedman:2003ax}) is equivalent to the Hamilton-Jacobi theory for the bulk equations of motion \cite{deBoer:1999xf}. It would be interesting to investigate if this is the case also for the first order flow for black hole solutions. In other words, it would be interesting to understand the Hamilton-Jacobi origin of the first order equations we have found in this paper. We leave these open questions to future investigations.

\section*{Acknowledgements}

We acknowledge clarifying and helpful correspondence with  Ioannis Papadimitriou. We would like to thank Gabriel Cardoso, Alejandra Castro, Gianguido Dall'Agata, Kristian Holsheimer, Ioannis Papadimitriou, Kostas Skenderis, Stefan Vandoren and Thomas Van Riet for interesting discussions.  We acknowledge support by the Netherlands Organization
for Scientific Research (NWO) under the VICI grant 680-47-603. This work is part of the D-ITP consortium, a program of the Netherlands
Organisation for Scientific Research (NWO) that is funded by the Dutch
Ministry of Education, Culture and Science (OCW).

\begin{appendix}

\section{Special geometry identities for the real submanifold \label{AppA}}

In case of zero axions $\mc N_{\Lambda \Sigma}=i\mc I_{\Lambda \Sigma}$.
The sections satisfy the relation
\begin{eqnarray}\label{uno}
<\mc V,\mc V\rq{}>=0=M_{\Lambda}(L^{\Lambda})\rq{}-L^{\Lambda}(M_{\Lambda})\rq{}\ .
\end{eqnarray}
However, from equation 4.35 and 4.38 of \cite{Andrianopoli:1996cm} we have
\begin{eqnarray}
L^{\Lambda}M_{\Lambda}=i(L^{\Lambda}\mc I_{\Lambda \Sigma}L^{\Sigma})=-\frac{i}{2}\ , \qquad \rar \qquad (L^{\Lambda}\mc I_{\Lambda \Sigma}L^{\Sigma})\rq{}=0
\nn
\end{eqnarray}
but that means
\begin{eqnarray}
M_{\Lambda}(L^{\Lambda})\rq{}+L^{\Lambda}(M_{\Lambda})\rq{}=0\ .
\end{eqnarray}
This, together with equation \eqref{uno} implies
\begin{eqnarray}\label{eq1}
L^{\Lambda}(M_{\Lambda})\rq{}=0 \ .
\end{eqnarray}
By definition\footnote{See eq. 4.35 of \cite{Andrianopoli:1996cm}. }, in absence of axions i.e. when $\re \mc N_{\Lambda \Sigma}=0$ we have
\begin{eqnarray}
M_{\Lambda}=i \mc I_{\Lambda \Sigma}L^{\Sigma}\ ,\qquad D_iM_{\Lambda}=-i\mc I_{ \Lambda \Sigma }D_iL^{\Sigma}\ ,
\nn
\end{eqnarray}
and, again, if we restrict to the real submanifold we have 
\begin{equation}
\mc{Q}=\frac{1}{2i}(dz^i \p_iK-d\bar z^{\bar\i} \p_{\bar\i}K)=0\ . \nn
\end{equation}
Then, 
\begin{eqnarray}
M_{\Lambda}'=\dot z^iD_iM_{\Lambda}+i\mc Q_r=(\textrm{zero axions})=\dot z^iD_iM_{\Lambda}=
-i\dot z^i\mc I_{ \Lambda \Sigma }D_iL^{\Sigma}=-i\mc I_{ \Lambda \Sigma }(L_{\Sigma})'
\nn
\end{eqnarray}
thus
\begin{eqnarray}
(\mc I_{\Lambda \Sigma}L^{\Sigma})'=-\mc I_{ \Lambda \Sigma }(L_{\Sigma})'\ .
\end{eqnarray}
This, together with \eqref{eq1}, imply also that
\begin{eqnarray}
L^{\Lambda} \mc I_{ \Lambda \Sigma }(L_{\Sigma})'=0\ .
\end{eqnarray}

The scalar fields dynamics for a spherically symmetric solution is described by a one dimensional system driven by an effective black hole potential \cite{Andrianopoli:2006ub}
\begin{eqnarray}
V_{BH}=-\frac12\mc Q^T\mc M(z,\bar z) \mc Q\ ,
\end{eqnarray}
in additon to the gauging scalar potential $V_g$. $\mc Q^T=(p^{\Lambda},q_{\Lambda})$ is the vector of charges and the symplectic matrix $\mc M$ is
\begin{eqnarray}
\mc M=\left(
\begin{array}
{cc} \mc I+\mc R\mc I^{-1}\mc R&-\mc R\mc I^{-1}\\
-\mc I^{-1}\mc R&\mc I^{-1}
\end{array}
\right)\ .
\end{eqnarray}
Black holes solutions with real scalars ($\mc R=0$) are supported by purely electric or purely magnetic charges, so the black hole potential in these cases is
\begin{eqnarray}
V_{BH}^{el}&=&-\frac12 q_{\Lambda}\mc I^{-1\, \Lambda \Sigma}q_{\Sigma}\ ,\qquad\quad
V_{BH}^{mag}=-\frac12 p^{\Lambda}\mc I_{ \Lambda \Sigma}p^{\Sigma}\ .
\end{eqnarray}

\section{Useful identities}

For the metric ansatz to be consistent with the scalar field dynamics the parameters satisfy that 
\begin{eqnarray}
a_0 a_1^3 =1 \ ,
\end{eqnarray}
for both electric and magnetic solutions, moreover
\begin{eqnarray}
z_\infty=\frac{a_0^{el}}{a_1^{el}}=\frac{a_1^{mag}}{a_0^{mag}}\ .
\end{eqnarray}

\begin{itemize}
\item The electric solution has
\begin{eqnarray}
L^{\Lambda}=e^{K/2}\frac{3^{3/4}\beta}{2\sqrt2}\left(
a_{\Lambda}+\frac{b_{\Lambda}}{r}
\right)\ ,
\end{eqnarray}
where $a_{\Lambda}=\{(\xi_0/\xi_1)^{3/4},(\xi_0/\xi_1)^{-1/4}\}$. Notice the relation
\begin{eqnarray}
\frac{3^{3/4}}{\sqrt2}a_{\Lambda}=\ell_{AdS}g_{\Lambda}\ .
\end{eqnarray}
\item The magnetic solution has
\begin{eqnarray}
L^{\Lambda}=e^{K/2}\frac{\beta}{2\sqrt2}\left(
a_{\Lambda}+\frac{b_{\Lambda}}{r}
\right)\ ,
\end{eqnarray}
where $a_{\Lambda}=\{(3\xi_0/\xi_1)^{-3/4},(3\xi_0/\xi_1)^{1/4}\}$. Notice the relation
\begin{eqnarray}
a_{\Lambda}=-\sqrt2\ell_{AdS}\mc G^{\Lambda}\ .
\end{eqnarray}
where $\mc G^{\Lambda}=(\mc I^{-1}_{\infty})^{\Lambda \Sigma}g_{\Sigma}$.
\end{itemize}
It follows that 
\begin{eqnarray}
(\mc I_{\infty})_{\Lambda \Sigma}\,a_{mag}^{\Sigma}=-3^{3/4} a_{\Lambda}^{el}\ .
\end{eqnarray}
The relation between parameters $Q_{\Lambda}$ and $a_{\Lambda},b_{\Lambda}$ is, for both solutions,
\begin{eqnarray}
Q_{\Lambda}=\frac{b_{\Lambda}}{a_{\Lambda}}\ .
\end{eqnarray}

\section{First order flow is sufficient to solve the second order equations of motion\label{Einsteinapp}}

In this appendix we show that the first order equations \eqref{zmagn} \eqref{defL}, supplemented by the hamiltonian constraint \eqref{VBHab} are sufficient to solve the full system of second order equations of motion. We show it explicitly for the magnetic squaring and the electric case can be worked out in full similarity.

Given that the Maxwell's and Bianchi equations are already solved by \eqref{formaa_metrica}- \eqref{ans_vec} and \eqref{sol-param1}, the equations left to verify are the Einstein's equations and the scalars second order equation. For spherically symmetric configurations just three of the Einstein's equations are nontrivial. Moreover, in the case of just one single scalar, it turns out that by solving the Einstein's equations the scalar equations of motion is automatically satisfied \cite{Toldo:2012ec}. Therefore we are left with these three equations to verify:
\begin{itemize}
\item First Einstein's equation (EQ1)
\begin{equation}\nn
\frac{2}{r} K' -\frac12 (K')^2 + K'' = 2 g_{zz} (\partial_r z)^2
\end{equation}
\item  Second Einstein's equation (EQ2)
\begin{equation}\nn
\mathbf{\widetilde{g}}^2 e^{-K} r^2 \left( 3-4 r  K' + r^2 (K')^2 - \frac12r^2 K''  \right) = - V_g
\end{equation}
\item  Third Einstein's equation (EQ3)
\begin{equation}\nn
-\frac{c_2}{r^2}-(2 \kappa r + c_1) K' -r (\kappa r + c_1 + \frac{c_2}{r}) K'' = \frac{p^{\Lambda} I_{\Lambda \Sigma} p^{\Sigma}}{h^2} 
\end{equation}
\end{itemize}
As anticipated, we will show now that these are satisfied given the first order flow \eqref{zmagn} \eqref{defL}, supplemented by the constraint \eqref{VBHab}.

\vspace{5mm}

\noindent We start from deriving with respect to $r$ eq. \eqref{defL}
\begin{equation}
L^{\Lambda} e^{-K/2} =  \tilde a^{\Lambda}+\frac{ \tilde b^{\Lambda}}{r}\,,
\end{equation}
obtaining
\begin{equation}\label{aa}
{L^{\Sigma}}' - \frac{ K'}{2} L^{\Sigma} +e^{K/2} \frac{b^{\Sigma}}{r^2} =0\,.
\end{equation}
We contract this with $i M_{\Sigma}$. Given that $L^{\Lambda} M_{\Lambda} = -i/2\,$,  we get
\begin{equation}\label{un}
(e^{-K/2})' = -i  2 \frac{ b^{\Lambda} M_{\Lambda} }{r^2}\,.
\end{equation}
As a further step we differentiate first eq \eqref{un}
\begin{equation}
 \left(- \frac{(K')^2}{2} + K''  \right) =4 i e^{K/2} \left( \frac{b^{\Lambda} M_{\Lambda}'}{r^2} - 2 \frac{b^{\Lambda} M_{\Lambda}}{r^3} \right)\,,
\end{equation}
so that EQ1 reads:
\begin{equation}
\frac{2}{r} K' + 4 i e^{K/2} \left( \frac{b^{\Lambda} M_{\Lambda}'}{r^2} - 2 \frac{b^{\Lambda} M_{\Lambda}}{r^3} \right) = 2 g_{zz} (\partial_r z)^2\,.
\end{equation}
At this point using the first order equation for $K'$ in \eqref{zmagn} and  the special K\"{a}hler identities in App. \ref{AppA} the right-hand side (RHS) of the previous equation reads
\begin{equation}
\text{RHS}= -4 {L^{\Lambda}}' I_{\Lambda \Sigma}  {L^{\Sigma}}' = -4 {L^{\Lambda}}' I_{\Lambda \Sigma}  \left( \frac{K'}{2} L^{\Sigma} - e^{K/2} \frac{b^{\Sigma}}{r^2} \right) = 4 {L^{\Lambda}}' I_{\Lambda \Sigma}  e^{K/2} \frac{b^{\Sigma}}{r^2} \,.
\end{equation}
We now massage the left hand side of EQ1 by making use of the special  K\"{a}hler relation $M_{\Lambda}' = -i I_{\Lambda \Sigma} (L^{\Sigma})'$ combined with eq. \eqref{aa}, namely
\begin{equation}\label{mlam}
M_{\Lambda}' = -i I_{\Lambda \Sigma} (L^{\Sigma})' = -i I_{\Lambda \Sigma} \left( \frac{K'}{2} L^{\Sigma} - e^{K/2} \frac{b^{\Sigma}}{r^2} \right)\,.
\end{equation}
The left hand side (LHS) turns out to be
\begin{equation}
\text{LHS}=4i e^{K/2} \frac{b^{\Lambda} M_{\Lambda}'}{r^2} = 4 e^{K/2} \frac{b^{\Lambda}}{r^2} I_{\Lambda \Sigma} (L^{\Sigma})' \,,
\end{equation}
so that we have proven that EQ1 is satisfied on the first order flow equations \eqref{zmagn} and  \eqref{defL}.

In order to verify the second Einstein's equation we use EQ1, which we have already verified. Plugging EQ1 in EQ2 we get
\begin{equation}
3(1-rK' +\frac14 r^2 (K')^2) - r^2 g_{zz} (\partial_r z)^2= -\frac{V_{g}}{\tilde{\xi}^2}e^{K}\,,
\end{equation}
which is satisfied too given the first order flow equations \eqref{zmagn} with a superpotential $\mathcal{W}$ such that
\begin{equation}
V_g = g^2 \left( g^{zz} \frac{\partial \mathcal{W}}{\partial z}\frac{\partial \mathcal{W}}{\partial z} - 3 \mathcal{W}^2 \right) .
\end{equation}
Finally, from eq. \eqref{un} we have
\begin{equation}
 - \frac{K''}{2} e^{-K/2}= \frac{(K')^2}{4} e^{-K/2}  -i  2 \frac{ b^{\Lambda} M_{\Lambda}' }{r^2}  +i  4 \frac{ b^{\Lambda} M_{\Lambda} }{r^3}\,,
\end{equation}
and also
\begin{equation}\label{quas}
 - \frac{K''}{2} =  e^{K/2} \left( - i (K') \frac{ b^{\Lambda} M_{\Lambda}}{r^2}  -i  2 \frac{ b^{\Lambda} M_{\Lambda}' }{r^2}  +i  4 \frac{ b^{\Lambda} M_{\Lambda} }{r^3} \right)\,.
\end{equation}
Using  \eqref{mlam} in \eqref{quas}  we come to the following useful expression:
\begin{equation}\label{due}
K'' + 2 \frac{K'}{r} = - 4 \frac{e^K}{r^4} b^{\Lambda} I_{\Lambda \Sigma} b^{\Sigma}\,.
\end{equation}
Using \eqref{due}, the constraint \eqref{VBHab} and the fact that $L^{\Lambda} I_{\Lambda \Sigma} L^{\Sigma} = -i/2 $, we get that the following equation holds
\begin{equation}
- \left( r^2 (\mathbb{\kappa}+\frac{c_1}{r} + \frac{c_2}{r^2}) K' \right)'  -\frac{c_2}{r^2}= e^K \frac{p^{\Lambda} I_{\Lambda \Sigma} p^{\Sigma}}{r^2}\,,
\end{equation}
that is precisely EQ3.

\vspace{3mm}
 The scalar (second order) equation of motion is automatically satisfied if the Einstein's equations are solved, so we showed that for our system the first order flow equations \eqref{zmagn}, \eqref{defL}, plus the constraint on the charges \eqref{VBHab} are sufficient.

\section{Computing the mass with the Ashtekar-Magnon-Das (AMD) prescription \label{AppADM}}

This is a recap of the main formulas of the mass computation for Anti-de Sitter black holes by means of the AMD procedure \cite{Ashtekar:1984zz,Ashtekar:1999jx}. The AMD techniques are valid for $d$-dimensional asymptotically AdS spacetime, but we restrict here our attention to four spacetime dimensions. 

The AMD procedure expresses the mass in terms of the integral of suitable contractions of the Weyl tensor over the conformal boundary at infinity. Since the black hole metric approaches asymptotically AdS, the integral is not divergent and well defined.

Details of the derivation of the can be found in the original papers \cite{Ashtekar:1984zz,Ashtekar:1999jx}, and for instance \cite{Chen:2005zj}. We give here a (very brief) summary of the formulas used and an explicit example for the computation of the mass. 

Given an asymptotically Anti-de Sitter configuration X with metric $g_{\mu \nu}$ with negative cosmological constant $\Lambda = -3/ l^2$, with a conformal boundary $\partial X$, one introduces a conformally rescaled metric $\overline{g}_{\mu \nu} = \Omega^2 g_{\mu \nu}$ such that on the conformal boundary $\partial X$ both $\Omega = 0$ and $ d \Omega \neq 0$ ($\Omega$ is defined up to a function $f$ that is nonzero on the boundary). As future reference, we will choose for our solutions $\Omega = \frac{l}{r}$.

If we denote as $\overline{C}^{\mu}_{\nu \rho \sigma}$ the Weyl tensor of the metric $g_{\mu \nu}$, with indices raised and lowered by the conformally rescaled metric $\overline{g}_{\mu \nu}$, and a vector $n_{\mu} = \partial_{\mu}\Omega$, one defines the quantity
\begin{eqnarray}\label{AMD1}
{\overline{E}^{\mu}}_{\nu} = l^2 \, \Omega \,\overline{n}^{\rho} \overline{n}^{\sigma} \overline{C}^{\mu}_{\nu \rho \sigma}\,.
\end{eqnarray}
The contraction of this quantity with an asymptotic Killing vector $K^{\mu}$ will give a conserved quantity, in this way:
\begin{eqnarray}\label{Qqq}
Q[K] = \frac{l}{8 \pi} \oint_{\Sigma} {\overline{E}^{\mu}}_{\nu} K^{\nu} d \overline{\Sigma}_{\mu} \,, 
\end{eqnarray}
Here $d\overline{\Sigma}_{\mu}$ is the area element of the spherical section of the conformal boundary. The authors of \cite{Ashtekar:1984zz,Ashtekar:1999jx} shown that $Q[K]$ is indeed a conserved charge, and this quantity does not depend on the conformal rescaling factor $\Omega$ defined before.

We are interested in the mass $M$ of the configuration, therefore we choose the time Killing vector $K = \partial / \partial_t$, therefore, from \eqref{Qqq} we have
\begin{eqnarray}\label{AMD2}
M = \frac{l}{8 \pi} \oint_{\Sigma} \overline{E}^t_{t} K^t d \overline{\Sigma}_t\,.
\end{eqnarray}

We show now how to compute the mass for the solutions described in sections \ref{ElApp} and \ref{MaApp}. In that case $l^2 = \frac{3 \sqrt3}{2 \sqrt{\xi_0 \xi_1^3}}$ and we take  $\Omega =l/ r $. The electric part of the Weyl tensor reads: 
\begin{eqnarray}
C^t_{rtr}  = -\frac{c_1}{ r^5}+O \left( \frac{1}{r^6} \right)\,.
\end{eqnarray}
Furthermore
\begin{eqnarray}
\overline{E}^t_t = \frac{l^2}{\Omega} \overline{g}^{\alpha r} \overline{g}^{\beta r} \overline{n}_r \overline{n}_r  C_{\alpha t \beta}^t = \frac{l^4}{r^4 \Omega^5} (g^{rr})^2 C_{rtr}^t\,,
\end{eqnarray}
so that the mass turns out to be:
\begin{eqnarray}
M = \frac{l}{8 \pi} \oint_{\Sigma} \overline{E}^t_{t} K^t d \overline{\Sigma}_t = - \frac{c_1}{2} \,.
\end{eqnarray}

\end{appendix}

\baselineskip 0 mm

\providecommand{\href}[2]{#2}\begingroup\raggedright\endgroup

\end{document}